\documentclass[11pt,a4paper,notoc]{article}
\usepackage{jheppub}
\usepackage{amsmath}         
\usepackage{amssymb}          
\usepackage{bbold}
\usepackage{ulem}
\usepackage{afterpage}
\usepackage{youngtab}
\usepackage{multirow}
\usepackage[capitalise]{cleveref}
\usepackage{slashed}
\usepackage{booktabs}
\usepackage{subcaption}
\usepackage{comment}
\usepackage{rotating}
\usepackage{url}
\usepackage{physics}
\usepackage{slashbox}
\usepackage{nicematrix}
\usepackage{xcolor}
\usepackage{color}
\usepackage{float}
\definecolor{linkcolor}{rgb}{0,0,0.6}
\definecolor{mygreen}{rgb}{0,0.6,0}
\definecolor{ballblue}{rgb}{0.13, 0.67, 0.8}

\crefname{section}{sec.}{secs.}
\crefname{table}{Tab.}{Tabs.}
\crefname{figure}{Fig.}{Figs.}
\crefname{equation}{Eq.}{Eqs.}
\crefname{appendix}{Appendix}{Appendix}

\newcommand{\SO}{\text{SO}}
\newcommand{\SU}{\text{SU}}
\newcommand{\U}{\text{U}}
\newcommand{\Sp}{\text{Sp}}

\def\beq{\begin{equation}}
\def\eeq{\end{equation}}

\def\gsim{\raise0.3ex\hbox{$\;>$\kern-0.75em\raise-1.1ex\hbox{$\sim\;$}}}
\def\lsim{\raise0.3ex\hbox{$\;<$\kern-0.75em\raise-1.1ex\hbox{$\sim\;$}}}

\usepackage[utf8]{inputenc}

\title{General vacuum stability of orbifold gauge breaking and application to asymptotic grand unification}

\author[a,b]{Giacomo~Cacciapaglia,}
\affiliation[a]{Universite Claude Bernard Lyon 1, CNRS/IN2P3, IP2I UMR 5822,  4 rue Enrico Fermi, F-69100 Villeurbanne, France}
\affiliation[b]{Quantum Theory Center (QTC) \& D-IAS, Southern Denmark Univ., Campusvej 55, 5230 Odense M, Denmark}
\emailAdd{g.cacciapaglia@ip2i.in2p3.fr}
\author[d]{Alan~S.~Cornell,}
\affiliation[d]{Department of Physics, University of Johannesburg,
PO Box 524, Auckland Park 2006, South Africa.}
\emailAdd{acornell@uj.ac.za}
\author[a,d]{Aldo~Deandrea,}
\emailAdd{deandrea@ip2i.in2p3.fr}
\author[a]{Wanda~Isnard,}
\emailAdd{wanda.isnard@ens-lyon.fr}
\author[c]{Roman Pasechnik,}
\emailAdd{roman.pasechnik@fysik.lu.se}
\author[c]{Anca~Preda,}
\emailAdd{anca.preda@fysik.lu.se}
\affiliation[c]{Department of Physics, Lund University, SE-223 62 Lund, Sweden}
\author[e]{Zhi-Wei Wang,}
\emailAdd{zhiwei.wang@uestc.edu.cn}
\affiliation[e]{School of Physics, The University of Electronic Science and Technology of China,\\
 88 Tian-run Road, Chengdu, China}

\begin{document}

\abstract{We examine the vacuum stability of gauge symmetry breaking in five dimensions, compactified on the $S_1/(\mathbb{Z}_2 \times \mathbb{Z}_2')$ orbifold. We consider $\SU(N)$, $\Sp(N)$, $\SO(2N)$ and $\SO(2N+1)$ theories in the bulk, and provide an exhaustive classification of possible parity assignments that lead to stable orbifolds and of the corresponding symmetry breaking patterns. We use these results in the search for viable asymptotic grand unification theories (aGUT), testing the stability criteria on models based on $\SU(6)$ and $\SU(8)$. As a result, we identify two viable aGUTs: a unique $\SU(6)$ pathway down to the Standard Model, and one $\SU(8)$ model leading to an intermediate Pati-Salam partial unification.}

\maketitle

%%%%%%%%%%%%%%%%%%%%%%%%%%%%%%%%%%%%%%%%%%%%%%%%%%%%%%%
\section{Introduction}
%%%%%%%%%%%%%%%%%%%%%%%%%%%%%%%%%%%%%%%%%%%%%%%%%%%%%%%

The idea that extra space dimensions may play a key role in describing gravitational and electromagnetic interactions dates back to the infancy of relativity and quantum theory \cite{Nordstrom:1914ejq,Kaluza:1921tu,Klein:1926tv}. Interest in extra dimensions has been revived by the development of string theory, where they are a necessary ingredient, and by the realisation that their characteristic size may be around the inverse TeV \cite{Antoniadis:1990ew} or even as large as the millimetre scale when only gravity can probe them \cite{Antoniadis:1998ig}. This leads to the definition of effective low energy theories described by a quantum field theory in $4+n$ dimensions, where the additional $n$ space-like dimensions must be compactified or have a non-trivial geometry \cite{Randall:1999vf} that hides their presence at large distances. In the past three decades, this realisation sparked the construction of a plethora of models, addressing various issues of the Standard Model (SM) of particle physics: from the origin of neutrino masses \cite{Dienes:1998sb,Grossman:1999ra} to the breaking of the electroweak symmetry with a Higgs boson \cite{Csaki:2002ur,Contino:2003ve,Scrucca_2004,Hosotani:2004wv,Hosotani:2005nz,Haba:2004bh} or without it \cite{Csaki:2003dt,Csaki:2003zu,Cacciapaglia:2004rb,Hosotani:2009jk}, and the construction of grand-unified theories (GUTs) \cite{Kawamura:1999nj,Hall:2001pg,Hebecker:2001jb,Dienes_2003,Kobayashi:2004ya}. Extra dimensions have also been used to explain geometrically the weakness of gravity as felt by the SM fields \cite{Arkani-Hamed:1998jmv,Randall:1999ee}, allowing for a lowering of the physical Planck scale (where quantum gravity effects should become relevant). While quantum field theories in extra dimensions are usually considered as effective theories, to be completed at energies not far from the compactification scale, five-dimensional constructions where the bulk couplings flow to a fixed point at high energy may lead to renormalisable models \cite{Gies:2003ic,Morris:2004mg}.

In all cases where gauge fields propagate in the bulk of the extra dimensions, a compactification based on an ``orbifold'' is required in order to consistently break the bulk gauge symmetry (when necessary) and ensure a chiral spectrum for the massless fermionic modes. An orbifold is defined as the covering of the extra space $\mathbb{R}^n$ by means of a set of discrete symmetries \cite{Hebecker:2001jb}, hence leading to a finite classification of all possibilities (see, for instance, Ref. \cite{Nilse:2007zz}). In this work, we focus on five dimensions, where the most general orbifold is based on two parity symmetries, $S^1/(\mathbb{Z}_2\times\mathbb{Z}_2^\prime)$, leading to the definition of an interval $[0, \pi R]$ with specific boundary conditions for the propagating fields. Phenomenological models have been considered both with flat space and with warped geometry \cite{Randall:1999ee}. 

An orbifold parity can only break the bulk gauge group $G$ to a maximal subgroup $H$, preserving the rank \cite{Hebecker:2001jb}.  Depending on the choice of parities and the symmetry breaking pattern, it may occur that a massless scalar mode is present in the spectrum, emerging from the 5D gauge multiplet. We will refer to this state as a gauge-scalar. This state is protected by the 5D gauge invariance, hence its potential is only generated at loop-level, {\it \`a la} Coleman-Weinberg, and proportional to the compactification scale $m_{\rm KK} = 1/R$ \cite{Antoniadis_2001,von_Gersdorff_2002}. The tantalising insensitivity to the cut-off of the theory \cite{HATANAKA_1998,Hosotani:2007kn} inspired the identification of such gauge-scalars with the Higgs fields in the SM, hence prompting the construction of gauge-Higgs unification models in flat \cite{Hosotani:2004wv} and warped space \cite{Hosotani:2005nz,Hosotani:2007kn}, with the latter being equivalent to holographic versions of composite (Goldstone) Higgs models \cite{Contino:2003ve}. Models of gauge-Higgs grand unification have also been studied \cite{Hosotani:2015hoa,Furui:2016owe,Maru:2019lit,Angelescu:2021nbp,Angelescu:2022obm}. A non-trivial vacuum expectation value (VEV) for the gauge scalar is, therefore, generated at one loop by the fields propagating in the bulk, {\it in primis} the gauge fields themselves. It further breaks $H$ to a smaller subgroup $H_v$, with reduced rank. It has been shown that the VEV can be mapped into a modification of the field boundary condition via a 5D gauge transformation \cite{Haba_20035D}. Hence, the symmetry breaking is due to a non-trivial Wilson line along the extra dimension \cite{Hosotani:1983xw,Hosotani:1988bm}. As the potential is periodic, the VEV acquires discrete values in units of the inverse compactification length-scale $R^{-1}$, such that small VEVs (preferred by the Higgs mechanism) are usually obtained via some form of fine tuning in the field content \cite{Cacciapaglia_2006} or couplings \cite{Scrucca_2003} in the theory.

It is well known that special values of the VEV exist, where the unbroken symmetry is enhanced to match that of an orbifold breaking. In other words, it may occur that one starts with an orbifold theory breaking the bulk gauge symmetry $G$ to $H$, while the minimum of the gauge-scalar potential corresponds to new boundary conditions that preserve a different maximal subgroup $H' \equiv H_v$. This fact would indicate an {\it orbifold instability} of the initial model $G/H$, which is better described by a different orbifold with a different symmetry breaking pattern $G/H'$. 

In this work, we systematically investigate the general conditions leading to the stability or instability of a gauge symmetry breaking orbifold in five dimensions, for bulk gauge groups $\SU(N)$, $\SO(N)$ and $\Sp(2N)$. For this purpose, we will consider the most minimal field content consisting of the gauge multiplet alone. We will find that not all symmetry breaking patterns are possible from stable orbifolds, hence providing significant model building constraints. We are particularly interested in applying these criteria to models of asymptotic grand unification (aGUT) \cite{Cacciapaglia:2020qky,Khojali:2022gcq,Cacciapaglia:2023ghp,Cacciapaglia:2023kyz}, where gauge couplings flow towards the same ultraviolet fixed point \cite{Bajc_2016}. In five dimensional models, the bulk fermion content is highly limited to small multiplicities and low-dimensional representations of the bulk gauge group, hence the gauge multiplet contribution to the gauge-scalar potential usually determines the vacuum of the theory. We will see this in practice within some concrete examples. We remark that unstable orbifolds may still be of phenomenological interest if bulk fermions and/or scalars overwhelm the gauge contribution to the potential, as may be the case for large multiplicities \cite{Haba:2004qf} or large gauge representations \cite{Cacciapaglia_2006}. Our results are of particular interest for minimal models of all types, and particularly for aGUTs, where the number and representation of bulk fermions are strongly constrained by the existence of fixed points \cite{Cacciapaglia:2023kyz}.

The paper is organised as follows: In section \ref{sec:methodology} we introduce the methodology, starting from the parity definitions and the general formulas for the effective potential. Specific subsections are dedicated to the relevant cases of $\SU(N)$, $\Sp(2N)$, $\SO(2N)$ and $\SO(2N+1)$. Section \ref{sec:agut} identifies minimal aGUT candidates that satisfy the orbifold stability criteria and studies their viability in more detail; by investigating the embedding of SM fermions, the Yukawa sector and the presence of UV fixed points. In section \ref{sec:summary} we give a summary of the results and our conclusions. A more detailed discussion on the parity definitions and the explicit computation of the effective potential can be found in the Appendices.

%%%%%%%%%%%%%%%%%%%%%%%%%%%%%%%%%%%%%%%%%%%%%%%%%%%%%%%
\section{Methodology and classification}
\label{sec:methodology}
%%%%%%%%%%%%%%%%%%%%%%%%%%%%%%%%%%%%%%%%%%%%%%%%%%%%%%%

We consider a five dimensional gauge theory where the extra spatial dimension $x^5$ is compactified on a flat $S^1/(\mathbb{Z}_2\times\mathbb{Z}_2')$ orbifold. This orbifold is defined as the covering of the infinite  one dimensional line $\mathbb{R}^1$ by two mirror parities $r_1$ and $r_2$, centred on the two boundaries of the interval, which we conventionally indicate as $x^5 = 0$ and $x^5 = \pi R$, respectively. On the 5D gauge fields, the parities act as follows:
\begin{equation} \label{eq:parityact}
    A_\mu (r_k(x^5)) = P_k \cdot A_\mu (x^5) \cdot P_k\,, \quad A_5 (r_k(x^5)) = - P_k \cdot A_5 (x^5) \cdot P_k\,,
\end{equation}
for $k=1,2$, and where $P_k$ are two independent parity matrices generated by the bulk gauge group $G$.
The two mirror parities commute, i.e.
\begin{equation}
    r_1\ast r_2 = r_2 \ast r_1\,,
\end{equation}
hence the parity matrices $P_k$ also commute. For any bulk gauge group $G$, therefore, the parity matrices $P_k$ are generated by elements of the Cartan subalgebra of $G$, while satisfying the condition $P_k^2 = \mathbb{1}$. In general, we can define
\begin{equation}
    \Omega (\theta_i) = \prod_{i=1}^{N_C} \exp (i \theta_i X_i^C)\,,
\end{equation}
where $X_i^C$ are the $N_C$ Cartan generators. Then, the most general parity matrix $P$ can be written as:
\begin{equation} \label{eq:generalP}
    P(\theta_0,\theta_i) = e^{i \theta_0} \Omega (\theta_i)
\end{equation}
for a choice of $\theta_i$, $i=0, \dots N_C$ that satisfies $P\cdot P=\mathbb{1}$. Note that the overall sign of $P$ is unphysical, c.f. Eq.~\eqref{eq:parityact}.

If at least one of the parity matrices is non-trivial, i.e. $P_k \neq \mathbb{1}$, the bulk gauge group is broken. More specifically, each parity breaks $G \to H_k$, while the unbroken 4D gauge symmetry is the intersection $H = H_1 \cap H_2$. In practice, the $G$ gauge multiplet can be decomposed under $H$, and each component is assigned a well defined parity $\pm 1$ under $r_1$ and $r_2$. Hence, the components can be classified in terms of a pair of parities, where conventionally we will list those of the vector $A_\mu$, as those for $A_5$ having opposite signs. The components with parities $(+,+)$ contain spin-1 zero modes, matching the unbroken generators of $H$. Instead, components with parities $(-,-)$ correspond to massless gauge-scalars, arising from $A_5$. Once the latter develops a VEV, the effective potential determining its value is computed by properly applying the one-loop Coleman-Weinberg formalism \cite{Antoniadis_2001}, which depends on the spectrum as a function of the VEVs. 

In general, the spectrum will have two kinds of contributions \cite{Cacciapaglia_2006}: 
\begin{eqnarray}
    m_{n,i} = \frac{n+a_i}{R} &\Rightarrow & \Delta V_{\rm eff} \propto \mathcal{F}^+ (a_i) = \frac{3}{2} \sum_{n=1}^\infty \frac{\cos 2 \pi n a_i}{n^5} \,, \label{ppmass} \\
    m_{n,j} = \frac{n+1/2+a_j}{R} &\Rightarrow & \Delta V_{\rm eff} \propto \mathcal{F}^- (a_j) = \frac{3}{2} \sum_{n=1}^\infty (-1^n) \frac{\cos 2 \pi n a_j}{n^5} \,,  
    \label{pmmass}
\end{eqnarray}
where $a_i$ and $a_j$ are the component-specific functions of the VEVs normalised to the inverse compactification scale $R^{-1}$. Contributions of type $\eqref{ppmass}$ stem from components with parity $(+,+)$ and $(-,-)$, while contributions of type \eqref{pmmass} stem from parities $(+,-)$ and $(-,+)$. The latter feature no zero modes. Hence, the most general potential can be written schematically as \cite{Haba_2004,Cacciapaglia_2006}
\begin{equation} \label{eq:genVeff}
    V_{\rm eff} = C \ \left[ - 3\ \mathcal{V}_{R_G} + 4 \sum_f \mathcal{V}_{R_f} - \sum_s x_s \mathcal{V}_{R_s} \right]\,,
\end{equation} 
where $R_G = Adj$ for the gauge contribution, and $f$ and $s$ count the fermion and scalar bulk fields in the representations $R_f$ and $R_s$, respectively, with $x_s = 1$ for real and $x_s=2$ for complex scalar fields. The overall normalisation factor reads
\begin{equation}
    C = \frac{1}{32 \pi^6 R^4}\,.
\end{equation}
The functions $\mathcal{V}_R$ contain the contribution of the modes in each representation $R$, giving
\begin{equation}
    \mathcal{V}_{R} = \sum_{i} \mathcal{F}^+(a_i) + \sum_j \mathcal{F}^-(a_j)\,,
\end{equation}
where $i$ and $j$ count components with parity $(\pm,\pm)$ and $(\pm,\mp)$, respectively.

The functions $\mathcal{F}^\pm$ have notable properties that will be important for the stability analyses. First of all, we find that
\begin{equation}
    \mathcal{F}^+ (a+1/2) = \mathcal{F}^- (a)\quad \mbox{and} \quad \mathcal{F}^- (a+1/2) = \mathcal{F}^+ (a)\,.
\end{equation}
Furthermore, $\mathcal{F}^-$ can be expressed in terms of $\mathcal{F}^+$ as follows:
\begin{equation}
    \mathcal{F}^- (a) = - \mathcal{F}^+ (a) + \frac{1}{16} \mathcal{F}^+ (2a)\,.
\end{equation}
Both functions are periodic, $\mathcal{F}^\pm (a+n) = \mathcal{F}^\pm (a)$, and acquire an extremal value, i.e. they have minima or maxima, at integer, $a=n$, or semi-integer, $a=1/2+n$, values of the argument. At such points, the unbroken gauge group is maximal and it corresponds to an orbifold breaking. The starting orbifold is considered {\it stable} if all VEVs occur at $a=0$, while it is {\it unstable} and needs to be replaced by a different orbifold if some VEVs correspond to $a=1/2$ (maximal VEV configuration).
Hence, it suffices to evaluate the potential at $a=0$ and $a=1/2$ and check which configuration gives the smallest value. At such points, we have
\begin{equation}
    \mathcal{F}^+(0) = \frac{3}{2} \zeta(5) > 0\,,
\end{equation}
while
\begin{equation}
    \mathcal{F}^- (0) = \mathcal{F}^+ (1/2) = - \frac{15}{16} \mathcal{F}^+(0)\,, \quad \mathcal{F}^- (1/2) = \mathcal{F}^+ (0)\,.
\end{equation}
From the signs above, one establishes that components with parities $(\pm,\pm)$ tend to stabilise the starting orbifold, while components with parities $(\pm, \mp)$ destabilise it. 

We now define the orbifold parities and classify the stability of symmetry breaking patterns for different bulk gauge groups: $\SU(N)$, $\Sp(2N)$, $\SO(2N)$ and $\SO(2N+1)$. Whenever massless gauge-scalar modes appear, we determine the conditions ensuring that the effective potential generated by the gauge multiplet prefers zero VEVs.  When the orbifold parities do not allow massless gauge-scalars to appear, the orbifold is automatically stable. We leave exceptional groups for further investigation.

%%%%%%%%%%%%%%%%%%%%%%%%%%%%%%%%%%%%%%%%%%%%%%%%%%%%%%%
\subsection{Special unitary groups: $\SU(N)$ case}
\label{sec:SU(N)}
%%%%%%%%%%%%%%%%%%%%%%%%%%%%%%%%%%%%%%%%%%%%%%%%%%%%%%%

We first consider a $\SU(N)$ gauge group in the bulk. The most general symmetry breaking patterns and gauge-scalar potentials have been studied in Ref.~\cite{Haba_2004}, here we re-derive the main results and use them to probe the stability condition for this class of orbifolds. 

Using the definition of the parity matrix in Eq.~\eqref{eq:generalP}, the most general $\SU(N)$ parity matrix is diagonal with $\pm 1$ entries:
\begin{equation}
        P_{\SU(N)} = {\rm diag}(\underbrace{1,\cdots,1}_{A},\underbrace{-1,\cdots,-1}_{N-A})
    \end{equation}
where $A \in [\![  1,N ]\!]$. More details on the derivation of this matrix can be found in Appendix \ref{ParityDefinition}. This parity breaks the gauge group in the bulk as it contains two sub-blocks with different signs, the first with $A$ diagonal entries and the second with $N-A$:
    \begin{equation}
        \SU(N) \to \SU(A) \times \SU(N-A) \times \U(1)\,,
    \end{equation}
while for $A=1$ and $A=N$, the group is unbroken.
    \paragraph{}
For the orbifold $S^1/(\mathbb{Z}_2\times\mathbb{Z}_2')$, the breaking pattern is determined by the relative alignment of the $\pm 1$ entries in the two parity matrices. The most general scenario entails four  different blocks with distinct combinations of parities \cite{Haba_2004}:
\begin{eqnarray}
P_1&=& {\rm diag}(+1,\cdots,+1,+1,\cdots,+1,-1,\cdots,-1,-1,\cdots,-1), \nonumber\\
P_2&=& {\rm diag}(\underbrace{+1,\cdots,+1}_{p},\underbrace{-1,\cdots,-1}_{q},\underbrace{+1,\cdots,+1}_{r},\underbrace{-1,\cdots,-1}_{s})\,,
\end{eqnarray}
where $p+q+r+s=N$, and each block dimension $p,q,r,s \in [\![  0,N ]\!] $. These parities leave the following 4D gauge group invariant:
\begin{equation}
    \SU(N) \to \frac{\U(p)\times \U(q) \times \U(r) \times \U(s)}{\U(1)}\,,
\end{equation}
where $\U(0) \equiv \emptyset$ and the $N$-dimensional identity matrix generator is removed (hence, maximally three $\U(1)$ factors can arise). Note that the overall sign of the parity matrix is irrelevant, as only relative signs matter. Hence, there are equivalences among various configurations of $(p,q,r,s)$, as follows:
\begin{eqnarray}
    \label{eq:parity equivalences1}
    P_1 \to - P_1 & \Rightarrow & p \leftrightarrow r,\;\; q \leftrightarrow s\,, \\
    \label{eq:parity equivalences2}
    P_2 \to - P_2 & \Rightarrow & p \leftrightarrow q,\;\; r \leftrightarrow s\,,
    \end{eqnarray}
    and flipping the sign of both parities ($i=1$ and $i=2$):
\begin{equation}
    \label{eq:parity equivalences3}
    P_i \to - P_i \Rightarrow  p \leftrightarrow s,\;\; q \leftrightarrow r\,. 
\end{equation}    
It is also equivalent (in flat space) to exchange the two parities:
\begin{equation}
    P_1 \leftrightarrow P_2\;\;  \Rightarrow \;\; q \leftrightarrow r\,. \\
\end{equation} 
Using these relations, one can show that any configuration can be recast into one where $p\geq r$ and $q\geq s$, and we will assume this condition in the rest of this section. The parities of the gauge fields read:
\begin{equation}
   (P_1,P_2) (A_\mu) = \begin{pNiceMatrix}[first-row,last-col=5,code-for-first-row=\scriptstyle,code-for-last-col=\scriptstyle] 
     p & q & r & s &  \\
      (+,+) &  (+,-) & (-,+) & (-,-) & p  \\
     (+,-) & (+,+) & (-,-) & (-,+) & q\\
     (-,+) & (-,-) & (+,+) & (+,-) & r  \\
     (-,-) & (-,+) & (+,-) & (+,+) & s  \\
    \end{pNiceMatrix} \;\; \equiv \; \Xi\,,
    \label{paritiesSUN}
\end{equation}
where the signs apply to the corresponding block.
The parities for $A_5$ can be easily deduced by flipping all the signs in Eq.~\eqref{paritiesSUN}. In general, two massless gauge-scalars are present, corresponding to the $(-,-)$ blocks, and they transform as bi-fundamental representations of two pairs of unbroken $\SU(K)$ factors:
\begin{equation}
    \varphi_{ps} = (F,1,1,\bar{F})\,, \qquad \varphi_{qr} = (1,F,\bar{F},1)\,.
\end{equation}
In the previous equation $F$ stands for the fundamental representation and the position in the list refers, respectively, to the $\SU(p)$, $\SU(q)$, $\SU(r)$, $\SU(s)$ subgroups. 
Gauge-scalar VEVs will further break the corresponding 4D gauge groups.

As already mentioned, there exist values of the VEVs that entail an enhanced symmetry, which corresponds to a rank-preserving orbifold breaking \cite{Haba_equivalentclasses}. For instance, if one component of the $\varphi_{ps}$ gauge-scalar is concerned, this is equivalent to flipping the parity signs in the corresponding positions of the parity matrices. Henceforth, a maximal VEV would correspond to a new parity configuration:
\begin{equation}
    \varphi_{ps} : \;\;\; (p,q,r,s) \to  (p-1, q+1, r+1, s-1)\,.
\end{equation}
Similarly, a maximal VEV for one component of $\varphi_{qr}$ corresponds to
\begin{equation}
    \varphi_{qr} : \;\;\; (p,q,r,s) \to  (p+1, q-1, r-1, s+1)\,.
\end{equation}
It is straightforward to show that, if one maximal VEV is preferred, then the largest number of VEVs in the gauge-scalar multiplet will also prefer to be maximal: this is $\text{min}(p,s) = s$ for $\varphi_{ps}$ and $\text{min}(q,r) = r$ for $\varphi_{qr}$.

Before studying the stability conditions, it is worth reminding the reader that, depending on the parity and their alignment, a 4D unbroken group can have various numbers of $\SU(K)$ factors and that a gauge-scalar is not always present. We distinguish the following cases:
\begin{enumerate}
    \item One-block case occurs when three blocks are trivial (e.g., $q=r=s=0$), hence $\SU(N)$ remains unbroken and no gauge-scalars arise.
    \item Two-block case occurs when two blocks are trivial, leading to
    \begin{equation}
        \SU(N) \to \SU(A) \times \SU(N-A) \times U(1)\,.
    \end{equation}
    A gauge scalar can be present (for instance, if $q=r=0$) or not (if $r=s=0$), depending on which blocks are trivial.
    \item Three-block case occurs when one block is trivial, and it always contains a gauge-scalar. Using the sign equivalences listed in Eq. \eqref{eq:parity equivalences1} to \eqref{eq:parity equivalences3}, the most general case can be parameterised by $r=0$, leading to
    \begin{equation}
        \SU(N) \to \SU(p) \times \SU(q) \times \SU(s) \times \U(1)^2\,,
    \end{equation}
    where, by convention, one gauge-scalar is in the bi-fundamental of $\SU(p) \times \SU(s)$ and $p \geq s$.
    \item Four-block case is the most general one with $p,q,r,s \geq 1$ and two gauge-scalars.
\end{enumerate}
In the following, we will always order the $\SU(K)$ factors according to the conventions established above.

For the stability analysis, we will consider the most general four-block case, where the other subcases can be obtained trivially by setting to zero the appropriate blocks.  As two gauge-scalars are present, the maximal VEV can occur in either one's direction, or in both, leading to the identification of three orbifolds that are connected to the original one:
\begin{itemize}
    \item If the maximal VEV is in $\varphi_{ps}$, the equivalent orbifold is given by the parities:
    \begin{align}
    &P_1={\rm diag}(+1\cdots,+1,+1,\cdots,+1,-1,\cdots,-1), \nonumber\\
    &P_2= {\rm diag}(\underbrace{+1,\cdots,+1}_{p-s},\underbrace{-1,\cdots,-1}_{q+s},\underbrace{+1,\cdots,+1}_{r+s})\,.
    \end{align}
    Hence, we end up with a three-block case:
    \begin{equation} \label{eq:varphipsmaxVEV}
    \SU(N) \to \SU(q+s) \times \SU(p-s) \times \SU(r+s) \times \U(1)^2\,,
    \end{equation}
    with a gauge-scalar in the bi-fundamental representation of $\SU(q+s)\times\SU(r+s)$.
    \item If the maximal VEV is in $\varphi_{qr}$, the equivalent orbifold is:
    \begin{align}
    &P_1= {\rm diag}(+1\cdots,+1,+1,\cdots,+1,-1,\cdots,-1), \nonumber\\
    &P_2= {\rm diag}(\underbrace{+1,\cdots,+1}_{p+r},\underbrace{-1,\cdots,-1}_{q-r},\underbrace{-1,\cdots,-1}_{r+s})\,,
    \end{align}
    also corresponding to a three-block case:
    \begin{equation}
    \SU(N) \to \SU(p+r) \times \SU(q-r) \times \SU(r+s) \times \U(1)^2\,,
    \end{equation}
    with gauge-scalar in the bi-fundamental representation of $\SU(p+r)\times\SU(r+s)$. Notably, a maximal VEV on this gauge-scalar connects this orbifold to the previous three-block one, and vice versa.
    \item If both gauge scalar VEVs are maximal, it yields:
    \begin{align}
    &P_1= {\rm diag}(+1\cdots,+1,+1,\cdots,+1,-1,\cdots,-1,-1\cdots,-1),\\
    \nonumber
    &P_2= {\rm diag}(\underbrace{+1,\cdots,+1}_{p-s+r},\underbrace{-1,\cdots,-1}_{q-r+s},\underbrace{+1,\cdots,+1}_{s},\underbrace{-1,\cdots,-1}_{r}).
    \end{align}
    We can check that this orbifold is connected to the previous three-block cases we saw appearing before, in a similar fashion.
\end{itemize}
\begin{figure}[H]
    \centering
    \includegraphics[height=4.5cm,width=10cm]{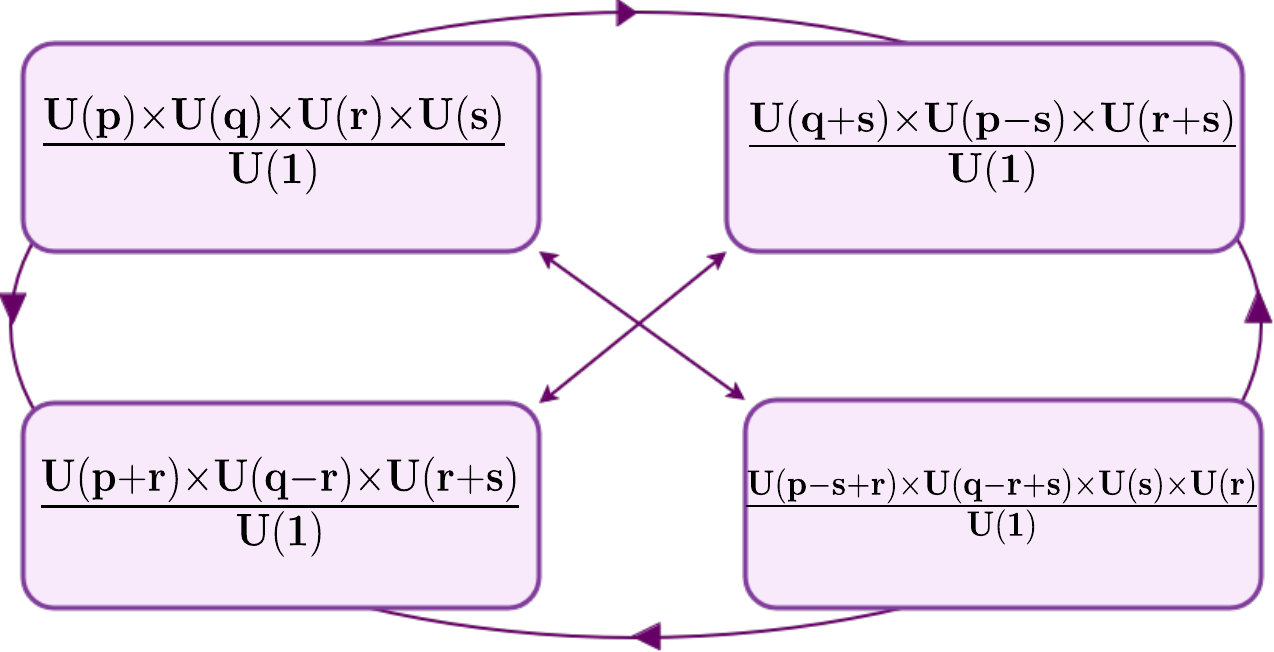}
    \caption{$\SU(N)$ orbifolds connected by maximal VEVs of the gauge-scalars. The four-block case in the bottom-right corner can reduce to a three-block one if $r=q+s$ or $s=p+r$, while the three-block case can reduce to a two-block one for $q=r$ or $p=s$. A stable orbifold is always preferred by the gauge-scalar potential. Note that the stability of three-block cases can be obtained by setting $r=0$ in the above diagram and considering only the first line.}
\label{fig:orbifoldschemeSUN}
\end{figure}
Henceforth, there are four orbifolds that are connected to each other by the generation of maximal VEVs for their gauge-scalars, as illustrated in Fig.~\ref{fig:orbifoldschemeSUN}. To check which one is preferred, we need to study the potential generated at one loop by the gauge multiplets in the bulk.

To compute the potential \cite{Cacciapaglia_2006,Haba_2004}, we normalise the gauge-scalar VEVs in such a way that the shifts $a_i$ in the spectrum always appear in the form of an integer times a parameter proportional to the VEV times the radius $R$. We call the latter $a$ as $\varphi_{ps}$ and $b$ as $\varphi_{qr}$, where we recall that all $s$ and $r$ VEVs are assumed to be equal. Up to the normalisation, see Eq.~\eqref{eq:genVeff}, the gauge potential reads \cite{Haba_2004}:
\begin{align} \label{eq:VeffSUN}
    \left. {\mathcal{V}_{Adj}}\right|_{\SU(N)} (a, b) = \left[ s^2 \mathcal{F}^+ (2 a)+ r^2 \mathcal{F}^+ (2b) + 2 r s \left( \mathcal{F}^- (a+b) + \mathcal{F}^- (a - b) \right) + \right.\nonumber \\
    \left.2 (p-s) \left( s \mathcal{F}^+ (a) + r \mathcal{F}^- (b) \right) + 2 (q-r) \left( r \mathcal{F}^+ (b) + s \mathcal{F}^- (a) \right) \right]\,,
\end{align}
where $V_{\rm eff} = - 3 C\ \mathcal{V}_{Adj}$. We then check the orbifold stability by computing the differences between the possible minima of the effective potential and the origin. We recall that there are three different possibilities: a maximal VEV in $\phi_{ps}$ such that $(a=1/2,b=0)$, in $\phi_{qr}$ leading to $(a=0,b=1/2)$, or both $(a=1/2,b=1/2)$:
\begin{eqnarray}
    -\Delta \mathcal{V}_{(1/2,0)} &=& \frac{93}{16} \zeta(5) s (2p-N)\,, \\
    -\Delta \mathcal{V}_{(0,1/2)} &=& \frac{93}{16} \zeta(5) r (2q-N)\,, \\
    -\Delta \mathcal{V}_{(1/2,1/2)} &=& \frac{93}{16} \zeta(5) (s-r) (p+r-q-s)\,.
\end{eqnarray}
The original orbifold is, therefore, stable if all $\Delta \mathcal{V} < 0$. For the first two, corresponding to the three-block orbifolds, at least one gives positive $\Delta \mathcal{V}$, signalling that four-block orbifolds are always unstable and must be described in terms of an orbifold with three or less blocks. 

The next step, therefore, is to check the stability of the three-block cases. Without loss of generality, we consider the general potential in Eq.~\eqref{eq:VeffSUN} with $r=0$. In which case
\begin{equation}
    -\Delta \mathcal{V}_{(1/2)} = -\Delta \mathcal{V}_{(1/2,0)} \propto s (2p-N)\,.
\end{equation}
Hence, the orbifold is stable only if $p \geq N/2$. For $p=N/2$, the two orbifolds are equivalent as they lead to the same symmetry breaking pattern. If the starting three-block orbifold is unstable, $p < N/2$, we can construct the one corresponding to the maximal VEV, which has $p'=q+s$, $q'=p-s$ and $s'=s$. We can check that the latter orbifold is stable as 
\begin{equation}
    p' =  N-p \geq  \frac{N}{2}\,.
\end{equation} 
We then notice that, among the two three-block orbifolds connected to a four-block one, one is always stable: they have $p'=q+s$ and $p'=p+r$, respectively, and as $N = p+r+q+s$, at least one of these two is greater than $ N/2$.

A two-block orbifold with gauge-scalars corresponds to $r=q=0$. The study of the minima of the potential is similar to the three-block case: $\Delta \mathcal{V}$ has the same expression as the three-block case but we always have $p>N/2$ as $p+s=N$.

In summary, we have shown that:
\begin{itemize}
    \item Four-block orbifolds are {\it always unstable}.
    \item Three-block orbifolds are {\it stable only if} $p\geq N/2$. The gauge scalar is always in the bi-fundamental representation of the largest subgroup, $\SU(p)$, and another $\SU(K)$ factor.
    \item Two-block orbifolds are {\it always stable}.
\end{itemize}

%%%%%%%%%%%%%%%%%%%%%%%%%%%%%%%%%%%%%%%%%%%%%%%%%%%%%%%
\subsection{Symplectic groups: $\Sp(2N)$ case}
%%%%%%%%%%%%%%%%%%%%%%%%%%%%%%%%%%%%%%%%%%%%%%%%%%%%%%%

Symplectic groups $\Sp(2N)$ are defined by a set of generators, being unitary matrices $X$ that satisfy:
\begin{equation}
    X \cdot E + E \cdot X^T = 0\,, \quad \mbox{with} \;\; E = \begin{pmatrix} 0 & \mathbb{1}_N \\ - \mathbb{1}_N & 0 \end{pmatrix}\,.
\end{equation}
The generators can be written in terms of $N\times N$ blocks as follows:
\begin{equation} \label{eq:Spblockform}
    X = \begin{pmatrix}
        A & B \\ C & -A^T \end{pmatrix}\,,
\end{equation}
with $B$ and $C$ being symmetric matrices, while $A$ is a unitary matrix.
Following this convention, the computation of the parity matrices can be found in Appendix~\ref{ParityDefinition}.
There are two inequivalent ways to define them, and we denote them by $P^{\rm I}$ and $P^{\rm II}$. They have the following expressions (in the same block form as the generators):
 \begin{equation}
        P^{\rm I}_{\Sp(2N)} = {\rm diag}(\underbrace{1,\cdots,1}_{A},\underbrace{-1,\cdots,-1}_{N-A})\otimes \begin{pmatrix} 1 & 0 \\ 0 & 1 \end{pmatrix}=P_{\SU(N)}\otimes \begin{pmatrix} 1 & 0 \\ 0 & 1 \end{pmatrix}\,,
    \end{equation}
\begin{equation} \label{eq:PIISp2N}
        P^{\rm II}_{\Sp(2N)} = {\rm diag}(\underbrace{1,\cdots,1}_{N})\otimes \begin{pmatrix} 1 & 0 \\ 0 & -1 \end{pmatrix}\,.
    \end{equation}
They lead to two different breaking patterns:
    \begin{equation}
    \begin{split}
       P^{\rm I}: & ~~\Sp(2N) \to \Sp(2A) \times \Sp(2(N-A)) \,, \\
       P^{\rm II}:& ~~ \Sp(2N) \to \SU(N) \times \U(1) \,.
    \end{split}
    \end{equation}
Having two types of parity matrices, the orbifold $S_1/(\mathbb{Z}_2\times\mathbb{Z}_2')$ features three general types of symmetry breaking patterns:
\begin{itemize}
    \item[I+I :] The first type is constructed out of two parities of type I, with the following general alignment:
    \begin{equation}
      \begin{array}{rcl} P_1&=&{\rm diag}(+1,\cdots,+1,+1,\cdots,+1,-1,\cdots,-1,-1,\cdots,-1)\otimes \begin{pmatrix} 1 & 0 \\ 0 & 1 \end{pmatrix} \,, \\
      \\
    P_2&=&{\rm diag}(\underbrace{+1,\cdots,+1}_{p},\underbrace{-1,\cdots,-1}_{q},\underbrace{+1,\cdots,+1}_{r},\underbrace{-1,\cdots,-1}_{s})  \otimes \begin{pmatrix} 1 & 0 \\ 0 & 1 \end{pmatrix}\,. \end{array}
    \end{equation}
    The parities of the adjoint components can be given in the block scheme of Eq.~\eqref{eq:Spblockform} in terms of the $\Xi$ pattern defined in Eq.~\eqref{paritiesSUN}:
    \begin{equation} \label{eq:P1P2Sp2N}
        (P_1,P_2)(A_\mu) = \left( \begin{array}{c|c} \Xi & \Xi \\ \hline \Xi & \Xi \end{array}\right)\,. 
    \end{equation}
    This leads to the following symmetry breaking pattern:
    \begin{equation}
        \Sp(2N) \to \Sp(2p) \times \Sp(2q) \times \Sp(2r) \times \Sp(2s)\,,  
    \end{equation}
    with two massless gauge scalars in the bi-fundamental representation,
    \begin{equation}
        \varphi_{ps} = (F, 1, 1, F)\,, \qquad \varphi_{qr} = (1,F,F,1)\,.
    \end{equation}
The potential for the gauge-scalar VEVs, computed in Appendix~\ref{Potential}, is the same as that of the $\SU(N)$ case in Eq.~\eqref{eq:VeffSUN}, up to a factor of 2. Therefore, we can immediately conclude that the vacuum structure will be analogous to that of the $\SU(N)$ case. The general conclusion is that the only stable orbifolds have either two blocks, or three blocks with $p\geq N/2$:
\begin{itemize}
    \item[A)] $\Sp(2N) \to \Sp(2p)\times \Sp(2(N-p))$, with or without gauge-scalar in $(F,F)$-representation of the unbroken gauge group,
    \item[B)] $\Sp(2N) \to \Sp(2p)\times \Sp(2q) \times \Sp(2(N-p-q))$, with $p\geq N/2$ and the gauge-scalar in $(F,1,F)$-representation of the unbroken gauge group.
\end{itemize}
    \item[II+II :] The second type is constructed out of two parities of type II. The misalignment occurs by exchanging the components with a $\pm 1$ in the blocks, leading to the most general scenario: 
    \begin{equation}
          \begin{array}{rcl} P_1&=& {\rm diag} (+1,\cdots,+1,+1,\cdots,+1)\otimes \begin{pmatrix} 1 & 0 \\ 0 & -1 \end{pmatrix} \,, \\
          \\
    P_2&=& {\rm diag}(\underbrace{+1,\cdots,+1}_{p},\underbrace{-1,\cdots,-1}_{q}) \otimes \begin{pmatrix} 1 & 0 \\ 0 & -1 \end{pmatrix}\,,
    \end{array}
    \end{equation}
    and leading to the parity scheme
    \begin{equation}
    \label{eq:A_II+II}
        (P_1,P_2)(A_\mu) = \left( \begin{array}{c|c} \begin{array}{cc} {(+,+)} & (+,-) \\ (+,-) & {(+,+)} \end{array} & \begin{array}{cc} {(-,-)} & (-,+) \\ (-,+) & {(-,-)} \end{array} \\ \hline \begin{array}{cc} {(-,-)} & (-,+) \\ (-,+) &{(-,-)} \end{array} & \begin{array}{cc} {(+,+)} & (+,-) \\ (+,-) & {(+,+)} \end{array} \end{array}\right)\,. 
    \end{equation}
    The definite parities apply to sub-blocks of size $p$ or $q$ (where $p+q=N$).
    The corresponding symmetry breaking pattern is
    \begin{equation}
        \Sp(2N) \to \U(p) \times \U(q)\,,
    \end{equation}
    with two gauge scalars transforming as the symmetric representation $S$ of the two gauge factors:
    \begin{equation}
        \varphi_{p} = (S,1)\,, \qquad \varphi_{q} = (1,S)\,.
    \end{equation}
    The computation of the effective potential can be found in Appendix~\ref{Potential}. Evaluating for equal values of all VEV components, up to the normalisation, the potential reads    
    \begin{equation}
       \left. \mathcal{V}^{\rm II+II}_{Adj}\right|_{\Sp(2N)} (a,b)= p^2 \mathcal{F}^+(2a) + q^2 \mathcal{F}^+(2b) + 2 pq \left( \mathcal{F}^-(a+b) + \mathcal{F}^-(a-b)\right)\,,
    \end{equation}
    where $a$ and $b$ stem from the symmetry of $\SU(p)$ and $\SU(q)$, respectively. As the potential difference is only given by $\mathcal{F}^-$ functions, we see that the global minimum is never at $(a,b)=(0,0)$ for any value of $p$ and $q$. Performing the gauge transformation to remove the maximal VEVs, we find that the only stable orbifold must have aligned parities $(P_1,P_1)$, leading to the only stable pattern:
    \begin{equation}
        \Sp(2N) \to \U(N)\,,
    \end{equation}
    without gauge-scalars.
    \item[II+I :] The third type has one parity of each type. We can always define the type-II parity as in Eq.~\eqref{eq:PIISp2N}, and misalign the parity of type-I, leading to the general case:
    \begin{equation}
          \begin{array}{rcl} P_1&=& {\rm diag} (+1,\cdots,+1,+1,\cdots,+1)\otimes \begin{pmatrix} 1 & 0 \\ 0 & -1 \end{pmatrix} \,, \\
          \\
    P_2&=& {\rm diag}(\underbrace{+1,\cdots,+1}_{p},\underbrace{-1,\cdots,-1}_{q}) \otimes \begin{pmatrix} 1 & 0 \\ 0 & 1 \end{pmatrix} \,.
    \end{array}
    \end{equation}
    The parity scheme on the adjoint components is
        \begin{equation}
        \label{eq:A_I+II}
        (P_1,P_2)(A_\mu) = \left( \begin{array}{c|c} \begin{array}{cc} {(+,+)} & (+,-) \\ (+,-) & {(+,+)} \end{array} & \begin{array}{cc} (-,+) & {(-,-)} \\ {(-,-)} & (-,+) \end{array} \\ \hline \begin{array}{cc} (-,+) & {(-,-)} \\ {(-,-)} & (-,+) \end{array} & \begin{array}{cc}{(+,+)} & (+,-) \\ (+,-) & {(+,+)} \end{array} \end{array}\right)\,,
    \end{equation}
    leading to the following symmetry breaking pattern:
    \begin{equation}
        \Sp(2N) \to \U(p) \times \U(q)\,.
    \end{equation}
    The massless gauge-scalar is in the bi-fundamental representation of $\SU(p) \times \SU(q)$:
    \begin{equation}
        \varphi_{pq} = (F,\bar{F})\,. 
    \end{equation}
    The potential, also computed in the Appendix~\ref{Potential}, up to the normalisation, reads:
    \begin{align}
      \left. V^{\rm I+II}_{Adj}\right|_{\Sp(2N)} (a) = \frac{1}{16} \left[ q^2 \mathcal{F}^+ (4a)  + 2 q (p-q) \mathcal{F}^+ (2a) \right]\,,
    \end{align}
    where we assumed $p>q$ without loss of generality. Containing only functions $\mathcal{F}^+$, it always has a minimum at zero VEV, hence, all such configurations are stable.
\end{itemize}

%%%%%%%%%%%%%%%%%%%%%%%%%%%%%%%%%%%%%%%%%%%%%%%%%%%%%%%%
\subsection{Special orthogonal groups: $\SO(2N)$ case}
\label{section:SO(2N)}
%%%%%%%%%%%%%%%%%%%%%%%%%%%%%%%%%%%%%%%%%%%%%%%%%%%%%%%%

Special orthogonal groups in even dimension share features similar to the symplectic case. The generators can be written in an analogous block form as follows:
\begin{equation}
    X = \begin{pmatrix} A & B \\ -B^T & D \end{pmatrix} \, ,
\end{equation}
where $A$ and $D$ are antisymmetric matrices and $B$ is a unitary matrix.
The general form of the parity matrices has been computed in Appendix~\ref{ParityDefinition}, and two types are present, $P^{\rm I}$ and $P^{\rm II}$. In block form, the parity matrices can be written in terms of the $\SU(N)$ parity matrix as follows:
\begin{equation}
    P_{\SO(2N)}^{\rm I} = P_{\SU(N)} \otimes \begin{pmatrix} 1 & 0 \\ 0 & 1 \end{pmatrix}  \,, \label{eq:PI_SO(2N)}
\end{equation}
\begin{equation}
    P_{\SO(2N)}^{\rm II} = P_{\SU(N)} \otimes \begin{pmatrix} 0 & -i \\ i & 0 \end{pmatrix} \,.
    \label{eq:PII_SO(2N)}
\end{equation}
They also lead to different breaking patterns:
\begin{equation}
    \begin{split}
       P^{\rm I}: & ~~ \SO(2N) \to \SO(2A) \times \SO(2(N-A)) \,, \\
       P^{\rm II}: & ~~ \SO(2N) \to \SU(N)\times \U(1)\,.
    \end{split}
\end{equation}
As before, the two parity types can be combined in three cases, leading to three different symmetry breaking patterns: 
\begin{itemize}
    \item[I+I :] If both parities are of type I, then the situation is very similar to the $\Sp(2N)$ case:
    \begin{equation}
        \begin{array}{rll} P_1&= &  {\rm diag}(+1,\cdots,+1,+1,\cdots,+1,-1,\cdots,-1,-1,\cdots,-1) \otimes \begin{pmatrix}  1 & 0 \\ 0 & 1 \end{pmatrix} \,,\\
    \\
        P_2&=& {\rm diag}(\underbrace{+1,\cdots,+1}_{p},\underbrace{-1,\cdots,-1}_{q},\underbrace{+1,\cdots,+1}_{r},\underbrace{-1,\cdots,-1}_{s}) \otimes\begin{pmatrix} 1 & 0 \\ 0 & 1 \end{pmatrix} \,.  \end{array}
    \end{equation}
    As a consequence, the adjoint components will receive the same block form parity scheme as in Eq.~\eqref{eq:P1P2Sp2N}, in terms of the $\Xi$ patterns of Eq.~\eqref{paritiesSUN}. This corresponds to the breaking pattern
    \begin{equation}
        \SO(2N) \to \SO(2p) \times \SO(2q) \times \SO(2r) \times \SO(2s)\,,  
    \end{equation}
    with two zero mode gauge-scalars:
    \begin{equation}
        \varphi_{ps} = (F, 1, 1, F)\,, \qquad \varphi_{qr} = (1,F,F,1)\,.
    \end{equation}
    The effective potential is computed in Appendix \ref{Potential} and, like for the symplectic case, it gives the same expression as in the $\SU(N)$ case up to a factor of 2. As a result, the vacuum structure will be the same as in the $\SU(N)$ case. The stable orbifolds are:
    \begin{itemize}
    \item[A)] $\SO(2N) \to \SO(2p)\times \SO(2(N-p))$ with or without gauge-scalar in $(F,F)$-representation of the unbroken gauge group,
    \item[B)] $\SO(2N) \to \SO(2p)\times \SO(2q) \times \SO(2(N-p-q))$ with $p\geq N/2$ and the gauge-scalar in $(F,1,F)$-representation of the unbroken gauge group.
    \end{itemize}
    \item[II+II :] This case differs slightly from the symplectic case due to the off-diagonal nature of the parity matrix in Eq.~\eqref{eq:PII_SO(2N)}. To identify generators with definite parities, we split the general form into two sets, as follows:
    \begin{equation}
        X=\underbrace{\begin{pmatrix}
            A_{+}/2&B_{s}\\
            -B_{s}&A_{+}/2
           \end{pmatrix}}_{X_{+}}
            +
        \underbrace{\begin{pmatrix}
             A_{-}/2&B_{a}\\
            B_{a}&-A_{-}/2   
    \end{pmatrix}}_{X_{-}} \,,
    \end{equation}
    where $A_\pm = A\pm D$ and the $B_{a/s}$ refer to the antisymmetric and symmetric parts of $B$, respectively. The parities of the $X_\pm$ generators are opposite, with the action of the parities on $X_+$ given by:
    \begin{equation}
        \label{eq:parities X+}
        \begin{array}{rll} P_1&= &{\rm diag}(+1,\cdots,+1,+1,\cdots,+1) \otimes \begin{pmatrix} 1 & 0 \\ 0 & 1 \end{pmatrix} \,, \\
        \\
        P_2&=&{\rm diag}(\underbrace{+1,\cdots,+1}_{p},\underbrace{-1,\cdots,-1}_{q}) \otimes \begin{pmatrix} 1 & 1 \\ 1 & 1 \end{pmatrix} \,.
        \end{array}
    \end{equation}
    Hence, the parity scheme on the components of the adjoint will be
    \begin{equation}
        (P_1,P_2)(A_\mu^+) = \left( \begin{array}{c|c} \begin{array}{cc} {(+,+)} & (+,-) \\ (+,-) & {(+,+)} \end{array} & \begin{array}{cc} {(+,+)} & (+,-) \\ (+,-) & {(+,+)} \end{array} \\ \hline \begin{array}{cc} {(+,+)} & (+,-) \\ (+,-) & {(+,+)} \end{array} & \begin{array}{cc} {(+,+)} & (+,-) \\ (+,-) & {(+,+)} \end{array} \end{array}\right)\,,
     \label{eq:A+ parities text}
    \end{equation}
    \begin{equation}
        (P_1,P_2)(A_\mu^-) = \left( \begin{array}{c|c} \begin{array}{cc} {(-,-)} & (-,+) \\ (-,+) & {(-,-)} \end{array} & \begin{array}{cc} {(-,-)} & (-,+) \\ (-,+) & {(-,-)} \end{array} \\ \hline \begin{array}{cc} {(-,-)} & (-,+) \\ (-,+) & {(-,-)} \end{array} & \begin{array}{cc} {(-,-)} & (-,+) \\ (-,+) & {(-,-)} \end{array} \end{array}\right)\,, 
         \label{eq:A- parities text}
    \end{equation}
    where $A^{\pm}_{\mu}$ corresponds to the generators $X_{\pm}$. The breaking pattern is 
    \begin{equation}
        \SO(2N) \to \U(p) \times \U(q)\,,
    \end{equation}
    with two gauge scalars transforming as an antisymmetric representation $A$ of the two gauge factors:
    \begin{equation}
        \varphi_{p} = (A,1)\,, \qquad \varphi_{q} = (1,A)\,.
    \end{equation}
    The potential for this scenario can be found in Appendix~\ref{Potential}, and it gives, up to the normalisation,
    \begin{equation}
        \left. \mathcal{V}_{Adj}^{\rm II+II}\right|_{\SO(2N)}  = \frac{3}{2} p^2 \mathcal{F}^+ (2 a) + \frac{3}{2} q^2 \mathcal{F}^+ (2b)+ 4 p q \left( \mathcal{F}^- (a+b) +\mathcal{F}^- (a-b) \right)\,.
    \end{equation}
    Similar to the symplectic case, the potential differences are only determined by $\mathcal{F}^-$ functions, hence the orbifold is never stable. The only stable case has $p=0$ or $q=0$, leading to aligned parities and the symmetry breaking pattern,
    \begin{equation}
        \SO(2N) \to \U(N)\,,
    \end{equation}
without gauge-scalars.
    \item[I+II :] This scenario is similar to the II+II combination discussed before, except that the $X_\pm$ generators have opposite parities only under the type-II parity. For $X_+$, we have the same parity definitions as in Eq.~\eqref{eq:parities X+}.
    Henceforth, while the gauge generators $A_+$ will have a parity scheme like in Eq.~\eqref{eq:A+ parities text}, for $A_-$ a sign needs to be flipped as compared to Eq.~\eqref{eq:A- parities text}. If we consider that the type-I parity is the second one, it leads to:
    \begin{equation}
        (P_1,P_2)(A^{-}_\mu) = \left( \begin{array}{c|c} \begin{array}{cc} {(-,+)} & (-,-) \\ (-,-) & {(-,+)} \end{array} & \begin{array}{cc} {(-,+)} & (-,-) \\ (-,-) & {(-,+)} \end{array} \\ \hline \begin{array}{cc} {(-,+)} & (-,-) \\ (-,-) & {(-,+)} \end{array} & \begin{array}{cc} {(-,+)} & (-,-) \\ (-,-) & {(-,+)} \end{array} \end{array}\right).
    \end{equation}
    The resulting breaking pattern is: 
    \begin{equation}
        \SO(2N) \to \U(p) \times \U(q)\,,
    \end{equation}
    with a massless gauge-scalar in the following representation:
    \begin{equation}
        \varphi_{pq} = (F,\bar{F})\,. 
    \end{equation}
    The potential for this case is computed in Appendix~\ref{Potential}, and it gives, up to the normalisation:
    \begin{equation}
        \left. \mathcal{V}_{Adj}^{\rm I+II}\right|_{\SO(2N)}(a) = \frac{q^2}{16} \mathcal{F}^{+}(4a) + \frac{q (p-q)}{8} \mathcal{F}^{+}(2a)+q^2 \mathcal{F}^{+}(2 a)\,. 
    \end{equation}
    As such, the minimum is always at $a=0$, thus making this scenario stable for all values of $p$ and $q$.
\end{itemize}

%%%%%%%%%%%%%%%%%%%%%%%%%%%%%%%%%%%%%%%%%%%%%%%%%%%%%%%%%%%
\subsection{Special orthogonal groups: $\SO(2N+1)$ case}
%%%%%%%%%%%%%%%%%%%%%%%%%%%%%%%%%%%%%%%%%%%%%%%%%%%%%%%%%%%

In the case of odd-dimensional $\SO(2N+1)$ groups, the Cartan algebra has the same number of elements as that of $\SO(2N)$. Due to the presence of an additional dimension, it is only possible to define a parity matrix of type-I, analogous to $P^{\rm I}_{\SO(2N)}$ in Eq.~\eqref{eq:PI_SO(2N)}, see Appendix \ref{ParityDefinition} for more details. Hence, the most general parity is a diagonal matrix with $2A+1$ plus signs and $2(N-A)$ minus signs, breaking the symmetry as
\begin{equation}
    \SO(2N+1) \to \SO(2A+1) \times \SO(2(N-A))\,.
\end{equation}
As a result, only one combination of parities can be formed: 
\begin{equation}
    \begin{array}{rll} P_1&=  & {\rm diag}(+1,\cdots,+1,+1,\cdots,+1,-1,\cdots,-1,-1,\cdots,-1) \,, \\
    P_2&=& {\rm diag}(\underbrace{+1,\cdots,+1}_{2p+1},\underbrace{-1,\cdots,-1}_{2q},\underbrace{+1,\cdots,+1}_{2r},\underbrace{-1,\cdots,-1}_{2s})\,, \end{array}
\end{equation}
which generates the symmetry breaking pattern
\begin{equation}
        \SO(2N+1) \to \SO(2p+1) \times \SO(2q) \times \SO(2r) \times \SO(2s)\,.  
\end{equation}
The parities of the gauge fields read
\begin{equation}
\label{eq:parities SO(2N+1)}
   (P_1,P_2)( A_\mu) = \begin{pNiceMatrix}[first-row,last-col=5,code-for-first-row=\scriptstyle,code-for-last-col=\scriptstyle] 
     2p+1 & 2q & 2r & 2s &  \\
      (+,+) &  (+,-) & (-,+) & (-,-) & 2p+1  \\
     (+,-) & (+,+) & (-,-) & (-,+) & 2q\\
     (-,+) & (-,-) & (+,+) & (+,-) & 2r  \\
     (-,-) & (-,+) & (+,-) & (+,+) & 2s  \\
    \end{pNiceMatrix} \,.
\end{equation}    
The two zero mode gauge-scalars are in the $ps$ and $qr$ sectors, such that
\begin{equation}
    \varphi_{ps} = (F, 1, 1, F)\,, \qquad \varphi_{qr} = (1,F,F,1)\,,
\end{equation}
where the notation is similar to the one used in the previous sections. From Eq.~\eqref{eq:parities SO(2N+1)}, we see that this case is very similar to the $\SO(2N)$ one with two type-I parity matrices, hence, we can apply the same results here.  

From the previous stability analysis, we see that the four-block case is always unstable. The only peculiarity of the $\SO(2N+1)$ case is that one of the block dimensions is always odd, while all the others are even. From the previous analysis, we can identify the following stable orbifolds:
\begin{itemize}
    \item[A)] $\SO(2N+1) \to \SO(2p+1) \times \SO(2q) \times \SO(2s)$ with $2p+1 > N$;
    \item[B)] $\SO(2N+1) \to \SO(2p) \times \SO(2q+1) \times \SO(2s)$ with $2p > N$;
    \item[C)] $\SO(2N+1) \to \SO(2p) \times \SO(2q) \times \SO(2s+1)$ with $2p > N$;
    \item[D)] All two block cases, with and without a gauge-scalar.
\end{itemize}
For the three-block cases, the gauge-scalar is a bi-fundamental representation of the first and the last $\SO(K)$ factors.

%%%%%%%%%%%%%%%%%%%%%%%%%%%%%%%%%%%%%%%%%%%%
\subsection{Summary of stability results}
%%%%%%%%%%%%%%%%%%%%%%%%%%%%%%%%%%%%%%%%%%%%

\begin{table}[H]
\centering
 \begin{tabular}{||c|c|c|c||} 
 \hline
 \textbf{Model}& \textbf{Breaking pattern} &\textbf{Stability} & \textbf{Gauge-scalar} \\ [1ex] 
 \hline\hline
\multirow{3}{*}{$\SU(N)$} & $\SU(A)\times \SU(N-A)\times \U(1)$ & $\forall A$ & $(F,\bar{F})$ or none  \\ [1ex] 
 \cline{2-4}
  & $\SU(p)\times \SU(q)\times \SU(s)\times\U(1)^2$&$p\geq N/2$ & $(F,1,\bar{F})$\\[1ex] 
 \hline
 \multirow{5}{*}{$\Sp(2N)$} & $\Sp(2A)\times \Sp(2(N-A))$&$\forall A$ & $(F,F)$ or none \\[1ex] 
 \cline{2-4}
 & $\Sp(2p)\times \Sp(2q)\times\Sp(2s)$&$p\geq N/2$ &$(F,1,F)$ \\[1ex] 
  \cline{2-4}
   & $\SU(A)\times\SU(N-A)\times \U(1)^2$&  $\forall A$ & $(F,\bar{F})$  \\ [1ex]
   \cline{2-4}
   & $\SU(N)\times\U(1)$&  always & none  \\ [1ex]
  \hline
  \multirow{5}{*}{$\SO(2N)$} & $\SO(2A)\times \SO(2(N-A))$&$\forall A$ & $(F,F)$ or none\\[1ex] 
 \cline{2-4}
   & $\SO(2p)\times \SO(2q)\times\SO(2s)$&$p\geq N/2$ & $(F,1,F)$ \\[1ex] 
  \cline{2-4}
 & $\SU(A)\times\SU(N-A)\times \U(1)^2$& $\forall A$ & $(F,\bar{F})$ \\[1ex] 
  \cline{2-4}
 & $\SU(N)\times\U(1)$& always & none \\[1ex] 
 \hline
 \multirow{4}{*}{$\SO(2N+1)$} & $\SU(2A+1)\times \SU(2(N-A))$ &$\forall A$ & $(F,F)$ or none \\ [1ex] 
 \cline{2-4}
  & $\SO(2p+1)\times \SO(2q)\times \SO(2s)$&$2p+1> N$ & $(F,1,F)$\\[1ex] 
 \cline{2-4}
  & $\SO(2p)\times \SO(2q+1)\times \SO(2s)$&$2p> N$ & $(F,1,F)$\\[1ex]
   \cline{2-4}
  & $\SO(2p)\times \SO(2q)\times \SO(2s+1)$&$2p> N$ & $(F,1,F)$\\[1ex]
  \hline
 \end{tabular}
 \caption{\label{tab:finalresults} List of stable orbifolds characterised by the symmetry breaking pattern and the presence of massless gauge-scalars. For the two-block cases, we consider $A \in [\![  0,N ]\!]$ such as we also include the one-block case and we also grouped together the orbifolds with and without gauge-scalar. For the three-block cases, we consider $p,q,s \in [\![  1,N-2 ]\!]$, with $p+q+s=N$.}
\end{table}

%%%%%%%%%%%%%%%%%%%%%%%%%%%%%%%%%%%%%%
\section{Minimal aGUT models}
\label{sec:agut}
%%%%%%%%%%%%%%%%%%%%%%%%%%%%%%%%%%%%%%

Models of asymptotic Grand Unification (or aGUTs) offer an alternative to the traditional unification paradigm. In their renormalisation group evolution, the couplings flow towards the same fixed point in the ultra-violet regime \cite{Bajc_2016}. Instead of acquiring the same value at a finite energy scale as in traditional GUTs, they only reach unification asymptotically and the unified symmetry is only restored approximately when the theory approaches the fixed point. Gauge theories in five dimensions may feature such ultra-violet fixed points, depending on the matter content of the model \cite{Morris:2004mg,Gies:2003ic}. The presence of a fixed point for the gauge couplings, as well as for the Yukawa ones, imposes severe limitations on the multiplicity and representation of the fermions in the bulk, and a general procedure for the identification of models has been established \cite{Cacciapaglia:2023kyz}.

In this work, we include the requirement that the 5D aGUT models should be based on stable orbifolds, as determined by the gauge-scalar potential from the gauge multiplet. In aGUTs, the fermions in the bulk are limited in number and can only appear in small-rank representations of the bulk gauge group, else the gauge fixed point is lost. As we will see in some examples below, their contribution to the gauge scalar potential can at most be comparable to that of the gauge multiplet, and they tend to move the minimum towards the breaking of the 4D gauge symmetry preserved by the orbifold. This feature could be useful in some cases. As a starter, we can inspect the orbifolds identified in Table~\ref{tab:finalresults} to find aGUT candidates. Exceptional groups can also be employed, with a candidate model identified in Ref. \cite{Cacciapaglia:2023ghp}, and we leave their systematic study for a future work.

In this work, we will consider two options regarding the unbroken 4D gauge group:
\begin{itemize}
    \item[A)] SM route, where the 4D unbroken group $H = \SU(3)_c \times \SU(2)_L \times U(1)_Y \times X$, with $X$ being an additional factorised (non-simple) group.
    \item[B)] Pati-Salam (PS)  \cite{PhysRevD.10.275} route, where $H = \SU(4)_{\rm PS} \times \SU(2)_L \times \SU(2)_R \times X$.
\end{itemize}
In principle, $X$ could be any group, however here we will only consider $X = \U(1)$ or $\U(1)^2$ for various reasons: non-Abelian groups are typically incompatible with the fermion embedding \cite{Cacciapaglia:2023kyz} and such a case would typically require a parity scheme with more than three blocks, which is unstable. The candidate models are listed in Table~\ref{tab:candidates}.
\begin{table}[htb]
\centering
 \begin{tabular}{||c|c|c|c|c||} 
 \hline
 \textbf{Model}& \textbf{Breaking pattern} &\textbf{Fermions} & \textbf{Fixed point} & \textbf{Gauge-scalar} \\ [1ex] 
 \hline\hline
 \multicolumn{5}{||c||}{SM route (A)} \\ \hline
$\SU(5)$ \cite{Cacciapaglia:2020qky} & $G_{\rm SM}$ & $\surd$ & \textcolor{red}{$\bf\times$} & none  \\ [1ex] 
  \hline
$\SU(6)$ (6A')\cite{Cacciapaglia:2023kyz} & $G_{\rm SM} \times \U(1)$ & $\surd$ & $n_g = 3$ & $(3,1)_{-1/3}$  \\ [1ex] 
  \hline
$\SU(6)$ \cite{Cacciapaglia:2023kyz} & $G_{\rm SM} \times \U(1)$ & \textcolor{red}{$\bf\times$} & $-$ & $(3,2)_{-5/3}$  \\ [1ex] 
  \hline
$\Sp(10)$  & $G_{\rm SM} \times \U(1)$ & \textcolor{red}{$\bf\times$} & $-$ & $(3,2)_{y}$  \\ [1ex] 
  \hline
$\SO(10)$ \cite{Haba_so10}  & $G_{\rm SM} \times \U(1)$ & \textcolor{red}{$\bf\times$} & $-$ & $(3,2)_{y}$  \\ [1ex] 
  \hline\hline
 \multicolumn{5}{||c||}{PS route (B)} \\ \hline
$\SU(8)$  & $G_{\rm PS} \times \U(1)^2$ & $\surd^\ast$ & $n_g \leq 3$ & $(4,1,2)$  \\ [1ex] 
  \hline
$\SO(10)$ \cite{Khojali:2022gcq}  & $G_{\rm PS}$ & $\surd$ & $2 \leq n_g \leq 5$ & none  \\ [1ex] 
  \hline
 \end{tabular}
 \caption{\label{tab:candidates} List of candidate aGUT models based on a stable orbifold. In the `Fermions' column we indicate if it is possible to embed the SM fermions as zero modes of a family of bulk fields, where the asterisk indicates the presence of additional non-chiral states. The column `Fixed point' indicates whether the bulk Yukawas feature a fixed point, and if it does we indicate $n_g$, the number of allowed bulk generations. We didn't compute the fixed point for models not featuring SM fermions as zero modes of bulk fields as they are already ruled out. Finally, the last column indicates the gauge-scalar quantum numbers under the SM or PS groups ($y$ indicates an undetermined hypercharge).}
\end{table}

A successful aGUT can only be constructed if the SM fermions can be embedded as zero modes of a family of bulk fermion fields with appropriate parity assignments. We consider the task achieved if the massless spectrum contains only the SM fermions and, at most, some additional non-chiral states. This criterion excludes the SM route models based on $\Sp(10)$ and $\SO(10)$ (see Appendix \ref{app:SO(10)}), as well as the second $\SU(6)$ candidate \cite{Cacciapaglia:2023kyz}. For $\SU(8)$, additional non-chiral zero modes are inevitable, similar to what we observed in the exceptional aGUT \cite{Cacciapaglia:2023ghp}. The fourth column indicates if the bulk Yukawas have a fixed point. This requirement usually imposes a bound on the number of bulk generations, and it allows us to exclude the $\SU(5)$ model \cite{Cacciapaglia:2020qky,Cacciapaglia:2023kyz}. We are therefore left with three potentially viable models, where the Yukawa analysis for the $\SU(8)$ one has been performed in this work. The $\SU(6)$ model has been identified in Ref. \cite{Cacciapaglia:2023kyz} and named as model 6A' (a less minimal version with the symmetric representation also exists, model 6S', however with a large number of massless bulk scalars). Note that $\SU(8)$ has previously only been considered in the supersymmetric case in six dimensions  \cite{Gogoladze:2003ci,Gogoladze:2005az}.

Before analysing the three remaining aGUTs, we want to recap how the ultra-violet fixed points are determined \cite{Cacciapaglia:2023kyz}, as this computation will be presented for the $\SU(8)$ model in the following subsections. At energies above the compactification scale $1/R$, where the 5D regime kicks in, any bulk coupling $h$ can be expressed in terms of an effective 't Hooft coupling, which takes into account the number of Kaluza-Klein (KK) modes below the renormalisation scale $\mu$:
\begin{equation}
    \tilde{\alpha}_h = \frac{h^2}{4 \pi} \mu R\,.
\end{equation}
For a gauge coupling $g$, the renormalisation group equation at one loop order reads
\begin{equation}
    2 \pi \ \frac{d \tilde{\alpha}_g}{d \ln \mu} = 2 \pi \ \tilde{\alpha}_g - b_5 \ \tilde{\alpha}_g^2\,,
\end{equation}
where $b_5$ is the 5D beta-function for gauge coupling evolution \cite{Cacciapaglia:2023kyz}. The fixed point exists if $b_5 > 0$, and it reads
\begin{equation}
    \tilde{\alpha}_g^\ast = \frac{2 \pi}{b_5}\,.
\end{equation}
For a model with a set of bulk Yukawa couplings $y$, one obtains coupled differential equations in the form
\begin{equation}
    2 \pi \ \frac{d \tilde{\alpha}_y}{d \ln \mu} = \left( 2\pi + \sum_{y'} c_{yy'} \tilde{\alpha}_{y'} - d_y \tilde{\alpha}_g\right) \tilde{\alpha}_y\,,
\end{equation}
where the coefficients $d_y$ and $c_{yy'}$ are computed at one loop order. In this case, the fixed points read
\begin{equation}
    \tilde{\alpha}_y^\ast = \sum_{y'} c_{yy'}^{-1} \left( d_y \tilde{\alpha}_g^\ast - 2 \pi \right)\,,
\end{equation}
where all values must be positive. We do not consider here the evolution of scalar quartic couplings.

In the remainder of this section, we will analyse the three models based on stable orbifolds, compute the potential for the gauge-scalars and their mass, and determine their viability. For $\SU(6)$ and $\SU(8)$, we will also provide a comparison with models based on unstable orbifolds.

%%%%%%%%%%%%%%%%%%%%%%%%%%%%%%%%%%%%%%%%%%%%%%%%%%%%%%%%%%%%%%%
\subsection{Stable $\SU(6) \to  \SU(3) \times \SU(2) \times \U(1)^2$ model}
%%%%%%%%%%%%%%%%%%%%%%%%%%%%%%%%%%%%%%%%%%%%%%%%%%%%%%%%%%%%%%%

The minimal stable $\SU(6)$ model in Table~\ref{tab:candidates} has been identified as model 6A' in Ref. \cite{Cacciapaglia:2023kyz}. The orbifold is based on the parities:
\begin{eqnarray}
    P_1={\rm diag}(+1\cdots,+1,+1,\cdots,+1,-1,\cdots,-1)\,,
    \nonumber \\
    P_2={\rm diag}(\underbrace{+1,\cdots,+1}_{p=3},\underbrace{-1,\cdots,-1}_{q=2},\underbrace{-1,\cdots,-1}_{s=1})\,,
\end{eqnarray}
hence leading to a stable three-block pattern. Contrary to the gauge multiplets, where the parities are uniquely determined by the parity matrices, for matter fields (i.e.~bulk fermions and scalars) an overall parity can be assigned, $\eta_{1,2} = \pm 1$. Hence, one can introduce fields in the same representation and with different parity assignments. However, relative parities of the components inside each field cannot be changed \cite{Cacciapaglia:2023kyz}. The SM fermions, therefore, can be obtained as zero modes of the following two fermion fields, both in the antisymmetric representation ${\bf 15}$ of $\SU(6)$:
\begin{eqnarray}
    \Psi_{\bf 15}^{(+,-)} &=& ({\bf 3,2})_{1/6,1}^{(+,+)} \oplus ({\bf 1,2})_{1/2,-2}^{(-,-)} \oplus ({\bf \bar{3},1})_{-2/3,1}^{(+,-)} \oplus ({\bf 1,1})_{1,1}^{(+,-)} \oplus ({\bf 3,1})_{-1/3,-2}^{(-,+)}\,, \\
    \Psi_{\overline{\bf 15}}^{(-,-)} &=& ({\bf 3,1})_{2/3,-1}^{(-,-)} \oplus ({\bf1,1})_{-1,-1}^{(-,-)} \oplus ({\bf \bar{3},1})_{1/3,2}^{(+,+)} \oplus ({\bf \bar{3},2})_{-1/6,-1}^{(-,+)} \oplus ({\bf 1,2})_{-1/2,2}^{(+,-)}\,, 
\end{eqnarray}
where the components are labelled in terms of their quantum numbers as follows:
\begin{equation}
    (\SU(3)_c, \SU(2)_L)_{\U(1)_Y, \U(1)_Z}\,,
\end{equation}
where $\U(1)_Z$ is the additional non-SM charge symmetry.
The parities $(\eta_1,\eta_2)$ are indicated on the $\Psi$ fields. For the components, we recall that  $(+,+)$ entails a left-handed zero mode, while $(-,-)$ entails a right-handed one. Hence, $\Psi_{\bf 15}^{(+,-)} \supset q_L + l_L^c$ and $\Psi_{\bf \bar{15}}^{(-,-)} \supset u_R + e_R + d_R^c$, where ${X}^c$ indicates the charge conjugate of a field $X$.
Since the gauge-scalar $\varphi_0 = ({\bf 3},{\bf 1})_{-1/3,3}$ is a colour triplet, it must not acquire a VEV. To generate SM fermion masses, a bulk scalar ${\bf 15}$ is added
    \begin{equation}
       \Phi_{\bf 15}^{(-,+)} = ({\bf 1,2})_{1/2,-2}^{(+,+)} \oplus ({\bf 3,2})_{1/6,1}^{(-,-)} \oplus  ({\bf \bar{3},1})_{-2/3,1}^{(-,+)} \oplus ({\bf 1,1})_{1,1}^{(-,+)} \oplus ({\bf 3,1})_{-1/3,-2}^{(+,-)}\,, 
    \end{equation}
which contains the SM Higgs doublet zero mode living in the $(+,+)$ component. To generate neutrino masses and introduce a viable Indalo dark matter candidate \cite{Cacciapaglia:2020qky}, two singlets shall also be added:
\begin{equation}
    \Psi_{\bf 1}^{(-,-)} = ({\bf 1}, {\bf 1})_{0,0}^{(-,-)}\,, \qquad \Psi_{\bf 1'}^{(+,-)} = ({\bf 1}, {\bf 1})_{0,0}^{(+,-)}\,.
\end{equation}
Hence, one can write bulk Yukawa couplings for up-type quarks and for the two singlets as follows:
\begin{equation}
    \mathcal{L}_{\rm Yuk} = - Y_u\ \overline{ \Psi}_{\bf \overline{15}} \Psi_{\bf 15} \Phi_{\bf 15} - Y_\nu\ \overline{ \Psi}_{\bf 15} \Psi_{\bf 1} \Phi_{\bf 15} - Y_\chi\ \overline{ \Psi}_{\bf \overline{15}} \Psi_{\bf 1'} \Phi_{\bf 15}\,.
\end{equation}
These bulk Yukawas give mass to the up-type fermions of the SM, while the down-type ones remain massless and must receive a mass from other mechanisms (for instance, via localised Yukawa couplings). The existence of fixed points for both gauge and Yukawa couplings requires $n_g = 3$ (see result for model 6A with the singlet Yukawas in Ref. \cite{Cacciapaglia:2023kyz}).

\begin{figure}[tbp]
    \centering
        \includegraphics[height=6cm]{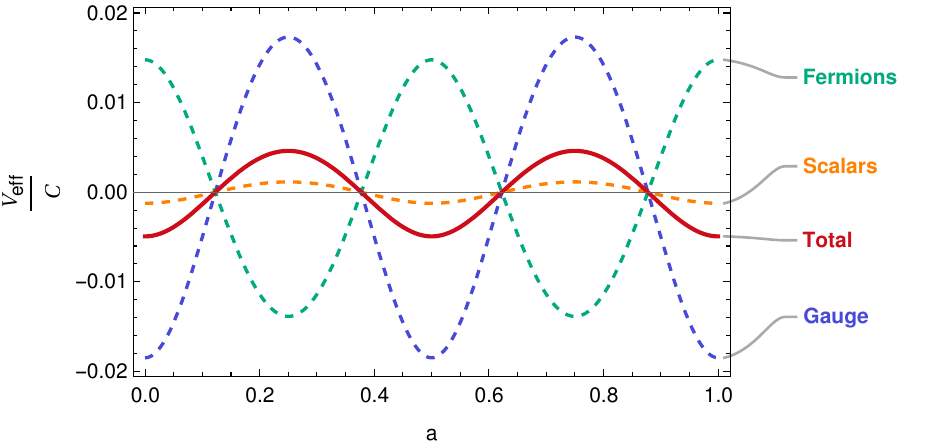}
        \caption{Plot of the gauge-scalar effective potential, and the individual contributions of the gauge, fermion and scalar sectors, for the model 6A'. The potential features two degenerate minima at $a=0$ and $a=1/2$, which leads to two equivalent descriptions of the same orbifold.}
        \label{fig:6A' plot}
\end{figure}

The total effective potential can be written in terms of the contribution of vector, fermion and scalar bulk fields as in Eq.~\eqref{eq:genVeff}.
The contributions of the adjoint and antisymmetric representations read
\begin{equation}
    \mathcal{V}_{Adj}(a) = \frac{5}{4} \mathcal{F}^+ (2a)\,, \qquad \mathcal{V}_{A}(a) = \frac{1}{8} \mathcal{F}^+ (2a)\,,
\end{equation}
respectively, where the latter does not depend on the parities. Hence,
\begin{equation} \label{eq:pot6A'}
    V_{\rm eff}(a) =  C \left(n_g - 4 \right) \mathcal{F}^+ (2a) = - C \ \mathcal{F}^+ (2a)\,,
\end{equation}
where we fixed $n_g = 3$. In general, the stability of the orbifold requires $n_g \leq 3$, consistently with the fixed points.
The profile of the potential for $n_g=3$ is plotted in Fig.~\ref{fig:6A' plot}, where we can see that the minimum remains at $a=0$, as expected. Notably, in this case the second minimum at maximal value $a=1/2$, degenerate with the $a=0$ one, corresponds to a completely equivalent orbifold. From Eq.~\eqref{eq:pot6A'}, we computed the mass of the gauge scalar at the minimum $a=0$, leading to
\begin{equation}
    m_\varphi^2 = \left. \frac{\partial^2}{\partial a^2} V_{\rm total} (a) \right|_{a=0}\, \frac{R^2}{2} = \frac{3}{16} \zeta(3) \frac{1}{\pi^4\ R^2}\,,
\end{equation}
where the factor $R^2/2$ comes from the relation between $a$ and the properly normalised gauge scalar field, $a = \frac{1}{\sqrt{2}} R \varphi_0$ \cite{Cacciapaglia_2006}. Via the bulk gauge interactions of $\Psi_{\bf 15}^{(+,-)}$, the gauge scalar couples to the quark and lepton doublets $\bar{q}_L l_L^c$, hence it will decay like a scalar leptoquark. Searches for leptoquarks at the Large Hadron Collider have been performed by both ATLAS \cite{ATLAS:2024fdw} and CMS \footnote{\href{https://twiki.cern.ch/twiki/bin/view/CMSPublic/SummaryPlotsEXO13TeV}{https://twiki.cern.ch/twiki/bin/view/CMSPublic/SummaryPlotsEXO13TeV}}, providing bounds on the masses in the ballpark of $2$~TeV for the QCD-induced pair-production channel. Hence, we can estimate a lower bound on the compactification scale in this $\SU(6)$ aGUT model,
\begin{equation}
    \frac{1}{R} = m_{\rm KK} \gtrsim 4 \pi^2 {\frac{1}{\sqrt{3\zeta(3)}}} \times 2~\mbox{TeV} \sim 50~\mbox{TeV}\,,
\end{equation}
which acts as a lower bound on the mass of the KK modes.

A variant of this model can be constructed by replacing $\Psi_{\bf 15}$ with the symmetric $\Psi_{\bf 21}$, at the price of embedding the Higgs boson into a bulk ${\bf 105}$ representation of $\SU(6)$, which contains multiple scalar zero modes. 

%%%%%%%%%%%%%%%%%%%%%%%%%%%%%%%%%%%%%%%%%%%%%%%%%%%%%%%%%%%%%%%
\subsection{Unstable $\SU(6) \to  \SU(3) \times \SU(2) \times \U(1)^2$ models}
%%%%%%%%%%%%%%%%%%%%%%%%%%%%%%%%%%%%%%%%%%%%%%%%%%%%%%%%%%%%%%%

For comparison, we also discuss the case of an unstable orbifold of $\SU(6)$, which was identified in Ref. \cite{Cacciapaglia:2023kyz} as the source of very appealing aGUT models, dubbed 6A, 6S and 6Aflip. In such models, one Higgs doublet arises as a gauge scalar, while a second one is in the bulk. 
The parity scheme is given by
\begin{eqnarray}
    P_1={\rm diag}(+1\cdots,+1,+1,\cdots,+1,-1,\cdots,-1)\,,\\
    \nonumber
    P_2={\rm diag}(\underbrace{+1,\cdots,+1}_{q=3},\underbrace{-1,\cdots,-1}_{p=2},\underbrace{+1,\cdots,+1}_{s=1})\,,
\end{eqnarray}
which is unstable as $p=2<3$. The model 6A has the same bulk representations as model 6A' up to different parities. In the model 6Aflip, the fermions are embedded in two fundamental and one 3-index antisymmetric representations of $\SU(6)$:
\begin{eqnarray}
    \Psi_{\bf 20}^{(-,-)}&=& ({\bf \bar{3},2})_{-1/6,3/2}^{(-,+)} \oplus ({\bf 3,1})_{2/3,3/2}^{(-,-)} \oplus ({\bf 1,1})_{-1,3/2}^{(-,-)} \oplus ({\bf 3,2})_{1/6,-3/2}^{(+,+)} \nonumber \\ & & \oplus ({\bf \bar{3},1})_{-2/3,-3/2}^{(+,-)}  \oplus ({\bf 1,1})_{1,-3/2}^{(+,-)} \,, \\
    \Psi_{\bf 6}^{(+,+)} &=& ({\bf 3,1})_{-1/3,1/2}^{(+,+)} \oplus ({\bf 1,2})_{1/2,1/2}^{(+,-)} \oplus ({\bf 1,1})_{0,-5/2}^{(-,+)} \,, \\
    \Psi_{\bf \bar{6}}^{(+,-)} &=& ({\bf \bar{3},1})_{1/3,-1/2}^{(+,-)} \oplus ({\bf 1,2})_{-1/2,-1/2}^{(+,+)} \oplus ({\bf 1,1})_{0,5/2}^{(-,-)} \,.
\end{eqnarray}
The bulk scalar field is the same in the two models and belongs to the antisymmetric $\bf 15$ representation.

\begin{figure}[H]
        \centering
        \includegraphics[height=6cm]{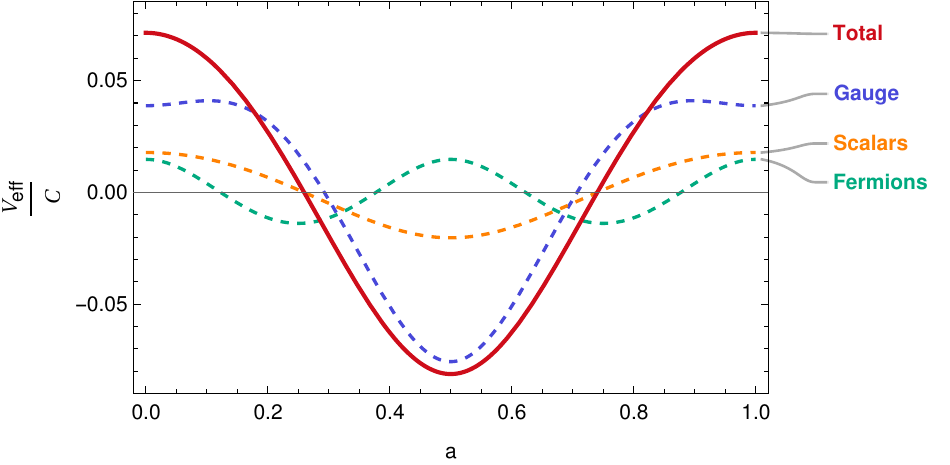}
        \caption{Effective potential for the models 6A and 6Aflip for $n_g=3$, based on an unstable orbifold. The global minimum is at $a=1/2$.}
        \label{fig:6A plot}
\end{figure}

As before, the effective potential for the gauge-scalar can be computed from contributions of relevant representations found as
\begin{eqnarray}
    \mathcal{V}_{Adj}(a) &=& \mathcal{F}^+ (2a) + 2 \mathcal{F}^+ (a) + 6 \mathcal{F}^- (a) \,, \\
    \mathcal{V}_{F}(a) &=&  \mathcal{F}^- (a)  \,, \\
    \mathcal{V}_{A}(a) &=&   \mathcal{F}^+ (a) + 3 \mathcal{F}^- (a) \,, \\
    \mathcal{V}_{A_3}(a) &=& 3 (\mathcal{F}^+ (a) + \mathcal{F}^- (a) )\,,
\end{eqnarray}
where the fermion and scalar contributions refer to fields with parities $(\pm,\pm)$. For parities $(\pm,\mp)$, it is enough to exchange $\mathcal{F}^+ \leftrightarrow \mathcal{F}^-$.
As model 6A contains a pair of $\Psi_{\bf 15}^{(+,-)}$ and $\Psi_{\bf \bar{15}}^{(-,-)}$ per generation, the fermionic contribution reads
\begin{equation}
    \mathcal{V}_{\rm fermions}^{\rm 6A}(a) = n_g \ \frac{1}{4} \mathcal{F}^+ (2a)\,.
\end{equation}
Model 6Aflip contains different matter content, consisting of two fundamental representations with flipped parity and a $3$ index antisymmetric representation $A_3$. The fermionic contributions to the effective potential can be computed in a similar way:
\begin{equation}
    \mathcal{V}_{\rm fermions}^{\rm 6Aflip}(a) =  n_g \ 4 (\mathcal{F}^+ (a) + \mathcal{F}^- (a)) = n_g \frac{1}{4} \mathcal{F}^+(2a)\,,
\end{equation}
which is accidentally the same as model 6A.
Hence, both models generate the same effective potential for the gauge-scalar, giving
\begin{equation}
    V_{\rm eff}(a) = C \left[ (n_g - 3) \mathcal{F}^+ (2a) - 8 (\mathcal{F}^+ (a) + 3 \mathcal{F}^- (a)) \right]\,.
\end{equation}
The profile of the effective potential for $n_g=3$ is plotted in Fig.~\ref{fig:6A plot}. We see that the gauge contribution determines the global minimum at $a=1/2$, which corresponds to an orbifold with a three-block symmetry breaking
\begin{equation}
    \SU(6) \to \SU(4) \times \U(1) \times \U(1)\,,
\end{equation}
incompatible with the SM. Reducing the number of bulk generations to $n_g = 1$ or $2$, as is possible in model 6Aflip, does not affect the global minimum, however, it will induce a metastable local minimum at $a=0$.

%%%%%%%%%%%%%%%%%%%%%%%%%%%%%%%%%%%%%%%%%%%%%%%%%%%%%%%%%%%%%
\subsection{$\SO(10)\to \SU(4)\times \SU(2)\times \SU(2)$}
%%%%%%%%%%%%%%%%%%%%%%%%%%%%%%%%%%%%%%%%%%%%%%%%%%%%%%%%%%%%%

This model has been studied in detail in Ref. \cite{Khojali:2022gcq}, and we will recap here the main features. The orbifold parity scheme reads
\begin{equation}
    \begin{array}{rll} P_1&= &{\rm diag}(+1,\cdots,+1,+1,\cdots,+1)\\
    P_2&=&{\rm diag}(\underbrace{+1,\cdots,+1}_{p},\underbrace{-1,\cdots,-1}_{q}) \end{array} \otimes \begin{pmatrix} 1 & 0 \\ 0 & 1 \end{pmatrix} \,,
\end{equation}
leading to the breaking pattern
\begin{equation}
    \SO(10) \to \SO(6) \times \SO(4) \equiv \SU(4) \times \SU(2) \times \SU(2)\,
\end{equation}
without a massless gauge-scalar. The SM fermions in PS transform as a left-handed $({\bf 4,2,1})$ and a right-handed $({\bf 4,1,2})$, which can be obtained as zero modes of a $\bf 16$ and $\bf \overline{16}$ in the bulk \cite{Khojali:2022gcq}. A bulk Yukawa, therefore, requires the addition of a real scalar $\bf 10$. Using the results in Ref. \cite{Cacciapaglia:2023kyz}, combined with the computations of the coupling running in Ref. \cite{Khojali:2022gcq}, we find the following renormalisation coefficients:
\begin{equation}
    b_5 = \frac{167-32 n_g}{6}\,, \qquad d_y = \frac{171}{8}\,, \qquad c_{yy} = 90\,.
\end{equation}
Hence, we determine the two fixed points to be
\begin{equation}
    \tilde{\alpha}^\ast_g = \frac{12 \pi}{167-32 n_g}\,, \qquad \tilde{\alpha}_y^\ast = \frac{(128 n_g - 155)\pi}{180 (167-32 n_g)}\,.
\end{equation}
We can deduce that a gauge fixed point exists for a number of generations up to $5$, and that the existence of a fixed point for the Yukawas excludes the case of $n_g=1$.  We are therefore left with $2 \leq n_g \leq 5$.

While this model is apparently viable, one issue highlighted in Ref. \cite{Khojali:2022gcq} is that, for three bulk generations, the value of the fixed point seems incompatible with the low energy value of the top Yukawa. Henceforth we deem the $\SO(10)$ aGUT to be disfavoured.

%%%%%%%%%%%%%%%%%%%%%%%%%%%%%%%%%%%%%%%%%%%%%%%%%%%%%%%%%%%
\subsection{Stable $\SU(8)\to \SU(4)\times \SU(2)\times \SU(2)\times \U(1)^2$ model}
%%%%%%%%%%%%%%%%%%%%%%%%%%%%%%%%%%%%%%%%%%%%%%%%%%%%%%%%%%%

A PS model can also be obtained from a bulk $\SU(8)$ via the following parity scheme:
\begin{eqnarray}
    P_1=(+1\cdots,+1,+1,\cdots,+1,-1,\cdots,-1) \,,\\
    \nonumber
    P_2=(\underbrace{+1,\cdots,+1}_{p=4},\underbrace{-1,\cdots,-1}_{q=2},\underbrace{-1,\cdots,-1}_{s=2})\,,
\end{eqnarray}
which also leaves two unbroken $\U(1)$ charges and features one massless gauge-scalar in the representation
\begin{equation}
    \varphi_0 = ({\bf 4,1,2})_{1,0} + ({\bf \bar{4},1,2})_{-1,0}\,,
\end{equation}
where the notation indicates the quantum numbers in the form 
\begin{equation}
(\SU(4),\SU(2),\SU(2))_{\U(1)_1, \U(1)_2} \,.
\end{equation}
It is convenient to identify the second $\SU(2)$ as $\SU(2)_R$, so that the gauge-scalar may be used to break the PS symmetry, if it develops a non-trivial VEV.

The SM fermions, transforming as left-handed $({\bf 4,2,1})$ and right-handed $({\bf 4,1,2})$ ones, can be minimally embedded in a two-index antisymmetric $\bf 28$ representation, which decomposes as 
\begin{equation}
\label{eq:parities of 28}
    \mathbf{28} = ({\bf 6,1,1})_{2,0}^{(+,+)}+ ({\bf 1,1,1})_{-2,2}^{(+,+)} + ({\bf 1,1,1})_{-2,-2}^{(+,+)} + ({\bf 4,2,1})_{0,1}^{(+,-)} + ({\bf 4,1,2})_{0,-1}^{(-,-)} + ({\bf 1,2,2})_{-2,0}^{(-,+)}\,. 
\end{equation}
For each component, we indicate the parities given by the parity matrix. Henceforth, the left-handed multiplet can be obtained from a $\Psi_{{\bf 28} {\rm L}}^{(+,-)}$, while the right-handed ones from $\Psi_{{\bf 28} {\rm R}}^{(+,+)}$. As one can deduce from Eq.~\eqref{eq:parities of 28}, both fields contain additional zero modes, which are in real or pseudo-real representations of the PS group, hence they can be given a mass via the breaking of the two $\U(1)$'s. This case is similar to what we encountered in the exceptional aGUT based on E$_6$ \cite{Cacciapaglia:2023ghp} and, while non-minimal, we will consider it as a viable model. 

The unified PS Yukawa can only be written down via a Higgs field embedded in the adjoint of $\SU(8)$, whose intrinsic parities are:
\begin{multline}
    \mathbf{63} = ({\bf 15,1,1})_{0,0}^{(+,+)} + ({\bf 1,3,1})_{0,0}^{(+,+)} + ({\bf 1,1,3})_{0,0}^{(+,+)} + ({\bf 1,1,1})_{0,0}^{(+,+)} + ({\bf 1,1,1})_{0,0}^{(+,+)} +\\
     \left[ ({\bf 4,2,1})_{1,-2}^{(+,-)} + ({\bf 4,1,2})_{1,0}^{(-,-)} + ({\bf 1,2,2})_{0,2}^{(-,+)} + \mbox{c.c.} \right]\,,
\end{multline}
where c.c.~indicates the conjugate representations with the same parities. Hence, we add a real bulk scalar field $\Phi_{\bf 63}^{(-,+)}$, which only contains a complex bi-doublet of $\SU(2)_L \times \SU(2)_R$ at the zero mode level. The bulk Yukawa will, therefore, be given by:
\begin{equation}
   \mathcal{L}_{\rm Yuk}\supset - Y_{\rm bulk} \overline{\Psi}_{{\bf 28} {\rm L}} \Phi_{\bf 63} \Psi_{{\bf 28} {\rm R}} + \mbox{h.c.}\,.
\end{equation}
To check if a fixed point for both gauge and Yukawa couplings exist, we computed the 5D running coefficients, obtaining
\begin{equation}
    b_5=\frac{8 (10-3 n_g)}{3}\,, \;\; d_y = \frac{157}{4}\,, \;\; c_{yy} = \frac{1}{8}\,.
\end{equation}
Hence, a UV fixed point for the gauge, $b_5>0$, exists only for $n_g \leq 3$. The same occurs for the Yukawa, leading to the fixed point values
\begin{equation}
    \tilde{\alpha}_g^\ast = \frac{3 \pi}{4 (10-3 n_g)}\,, \qquad \tilde{\alpha}_y^\ast = \frac{(96 n_g+151) \pi}{2 (10-3 n_g)}\,.
\end{equation}
Note that for $n_g = 3$, the Yukawa fixed point is non-perturbative, so in a realistic model one may need to localise one or two generations to the orbifold boundary.

\begin{figure}[H]
    \centering
    \includegraphics[height=6cm]{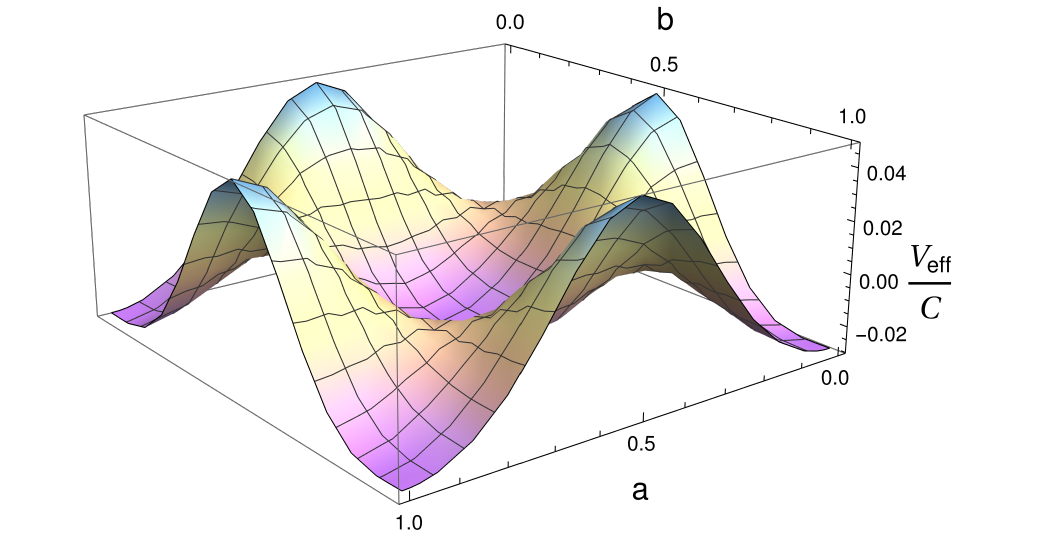}
    \caption{Plot of the gauge scalar effective potential for the stable $\SU(8)$ model with $n_g=3$. The potential features degenerate minima at $a=b=0$ and $a=b=1/2$, which lead to two equivalent descriptions of the same orbifold.}
    \label{fig:SU8}
\end{figure}

Finally, we computed the effective potential for the gauge-scalar. Being in a $({\bf 4,1,2})$ representation of the PS gauge group, it can have two independent VEVs, which we label $a$ and $b$. The contribution of the relevant representations gives
\begin{eqnarray}
    \mathcal{V}_{Adj}(a,b) &=& \frac{5}{4} \left(\mathcal{F}^+ (2a)+ \mathcal{F}^+ (2b) \right) + 2 \left( \mathcal{F}^+ (a+b) + \mathcal{F}^+ (a-b)\right)\,, \\
    \mathcal{V}_{\rm fermions}(a,b) &=& \frac{1}{4} \left( \mathcal{F}^+ (2a) + \mathcal{F}^+ (2b)\right) + \frac{1}{16} \left( \mathcal{F}^+ (2a+2b) + \mathcal{F}^+ (2a-2b)\right)\,, \\
    \mathcal{V}_{\rm scalar}(a,b) &=&  + \frac{1}{4} \left( \mathcal{F}^+ (2a) + \mathcal{F}^+ (2b) \right) + \mathcal{F}^- (2a) + \mathcal{F}^- (2b) +\nonumber \\
    && + 2 \left( \mathcal{F}^- (a+b) + \mathcal{F}^- (a-b) \right)\,,
\end{eqnarray}
where we have included the two $\bf 28$ representations for the fermion contribution, and the adjoint with appropriate parities for the scalar one. The total potential from Eq.~\eqref{eq:genVeff} reads, for $n_g=3$:
\begin{multline}
    V_{\rm eff}(a,b) = - C \left[ 4 \left( \mathcal{F}^+ (a+b) + \mathcal{F}^+ (a-b) \right) + \frac{1}{16} \left( \mathcal{F}^+ (4a) + \mathcal{F}^+ (4b)\right)\right. \\
    \left.- \frac{5}{8} \left(\mathcal{F}^+ (2a+2b) + \mathcal{F}^+ (2a-2b) \right)\right]\,.
\end{multline}
As shown in Fig.~\ref{fig:SU8}, this potential has two degenerate minima for $a=b=0$ and $a=b=1/2$, which correspond to the same orbifold and are also preferred by the gauge contribution to the potential. The same structure remains for fewer bulk generations. This result confirms the stability analysis from the previous section. 

%%%%%%%%%%%%%%%%%%%%%%%%%%%%%%%%%%%%%%%%%%%%%%%%%%%%%%%%%%%%%%%
\subsection{Unstable  $\SU(8)\to \SU(4)\times \SU(2)\times \SU(2)\times \U(1)^2$ model}
%%%%%%%%%%%%%%%%%%%%%%%%%%%%%%%%%%%%%%%%%%%%%%%%%%%%%%%%%%%%%%%

An interesting aGUT PS model can also be obtained by changing the parities to
\begin{eqnarray}
    P_1=(+1\cdots,+1,+1,\cdots,+1,-1,\cdots,-1),\\
    \nonumber
    P_2=(\underbrace{+1,\cdots,+1}_{p=4},\underbrace{-1,\cdots,-1}_{q=2},\underbrace{+1,\cdots,+1}_{r=2})\,,
\end{eqnarray}
which leads to an unstable orbifold, with a gauge-scalar transforming as $({\bf 1,2,2})$, i.e. the SM Higgs doublet.
In fact, the intrinsic parities of the adjoint read:
\begin{multline}
    \mathbf{63} = ({\bf 15,1,1})_{0,0}^{(+,+)} + ({\bf 1,3,1})_{0,0}^{(+,+)} + ({\bf 1,1,3})_{0,0}^{(+,+)} + ({\bf 1,1,1})_{0,0}^{(+,+)} + ({\bf 1,1,1})_{0,0}^{(+,+)} +\\
     \left[ ({\bf 4,2,1})_{1,-2}^{(+,-)} + ({\bf 4,1,2})_{1,0}^{(-,+)} + ({\bf 1,2,2})_{0,2}^{(-,-)} + \mbox{c.c.} \right]\,.
\end{multline}
As the intrinsic parities of the antisymmetric $\bf 28$ representation now read
\begin{equation}
    \mathbf{28} = ({\bf 6,1,1})_{2,0}^{(+,+)}+ ({\bf 1,1,1})_{-2,2}^{(+,+)} + ({\bf 1,1,1})_{-2,-2}^{(+,+)} + ({\bf 4,2,1})_{0,1}^{(+,-)} + ({\bf 4,1,2})_{0,-1}^{(-,+)} + ({\bf 1,2,2})_{-2,0}^{(-,-)}\,,
\end{equation}
a SM generation can be obtained from a single bulk field $\Psi_{\bf 28}^{(+,-)}$ without any extra zero modes. Furthermore, all Yukawa couplings stem from the gauge interactions, guaranteeing the presence of an attractive fixed point for both gauge and Yukawa couplings \cite{Cacciapaglia:2023kyz}. The gauge coupling running has
\begin{equation}
    b_5 = 28 - 4 n_g\,.
\end{equation}
Hence a fixed point exists for $n_g \leq 6$.

The simplest version of this model, however, suffers from the instability issue. The contribution of the gauge and one fermion to the gauge-scalar potential reads:
\begin{eqnarray}
    \mathcal{V}_{Adj}(a,b) &=& \mathcal{F}^+ (2a) + \mathcal{F}^+ (2b) + 2 \mathcal{F}^+ (a+b) + 2 \mathcal{F}^+ (a-b) \nonumber \\
    && + 8 \left(\mathcal{F}^- (a) + \mathcal{F}^- (b) \right)\,, \\
    \mathcal{V}_{28, (+,-)}(a,b) &=& \mathcal{F}^- (a+b) + \mathcal{F}^- (a-b) + 4 \left( \mathcal{F}^+ (a) + \mathcal{F}^+ (b) \right)\,, 
\end{eqnarray}
where $a$ and $b$ are the normalised VEVs of the two doublets in the gauge-scalar. For a number of bulk generations satisfying $n_g \leq 6$, the potential has a minimum at the value preferred by the gauge contribution, $a=b=1/2$, which corresponds to a different orbifold with breaking
\begin{equation}
    \SU(8) \to \SU(6) \times \SU(2) \times \U(1)\,,
\end{equation}
with gauge-scalar in the bi-fundamental representation.
This conclusion may be changed once the breaking of the PS group is included, which could occur at a scale much larger than $m_{KK}$. Hence this model remains of interest while requiring some engineering to be feasible and realistic.

Another possibility arises by adding supersymmetry. The main advantage is that the loop-induced potential for the gauge-scalar vanishes, and it can only be generated by supersymmetry breaking terms. Furthermore, the running of the gauge coupling is also modified \cite{Cacciapaglia:2023kyz}, leading to
\begin{equation}
    b_5^{\rm susy} = \frac{\pi}{2} (8 - 3 n_g)\,.
\end{equation}
This allows a fixed point for up to 2 bulk families (at least one must remain localised at one boundary of the orbifold). We leave a detailed investigation of this model for future studies.

%%%%%%%%%%%%%%%%%%%%%%%%%%%%%%%%%%%%%%%%%%%%%
\section{Summary and conclusions}
\label{sec:summary}
%%%%%%%%%%%%%%%%%%%%%%%%%%%%%%%%%%%%%%%%%%%%%

Orbifolds can be used to efficiently reduce extra-dimensional theories to four-dimensional ones that resemble the Standard Model, in particular, achieving a chiral spectrum of massless fermions and breaking the large gauge symmetry groups. However, the presence of massless scalar modes from the gauge multiplet may signal an instability in the orbifold configuration. We systematically studied the stability of orbifolds based on the most general (flat) interval $S^1/(\mathbb{Z}_2\times\mathbb{Z}_2')$, which breaks gauge groups $\SU(N)$, $\Sp(2N)$ and $\SO(N)$. The summary of stable configurations and symmetry breaking patterns is presented in Table~\ref{tab:finalresults}. In particular, we find that the bulk group can only be broken to, at most, three non-Abelian subgroups. The classification we present here can be used as a starting point for the construction of models in five dimensions, especially for scenarios with a limited number of bulk matter fields, which could act as stabilising agents.

We apply our results to the case of asymptotic Grand Unification models, where the fermion content in the bulk is limited by the requirement of fixed points for the gauge and Yukawa couplings. Henceforth, we have shown that the stability criterion is an effective way to skim off unrealistic models, and it complements the classification procedure delineated in Ref.~\cite{Cacciapaglia:2023kyz}. As a consequence, from all possible aGUTs based on $\SU(N)$, $\Sp(2N)$ or $\SO(N)$, we identify only two feasible minimal models:
\begin{itemize}
    \item[A)] An $\SU(6)$ model leading to the Standard Model at low energies;
    \item[B)] An $\SU(8)$ model leading to Pati-Salam at low energies.
\end{itemize}
Both models are based on a stable orbifold, and feature fixed points for both gauge and bulk Yukawa couplings. For the $\SU(6)$ model, the number of bulk generations is fixed to be 3, while the $\SU(8)$ model allows up to 3 generations in the bulk. An $\SO(10)$ model leading to PS also seems viable, however, the Yukawa fixed point predicted for three generations is at odds with the low energy value of the top Yukawa. Exceptional groups will be studied in a forthcoming publication.

Our results show that constraining requirements allow us to pin down {\it unique} candidates for aGUT models. Note, however, that we based our analysis on minimality. Some models based on unstable orbifolds may be salvaged if the potential for the massless gauge-scalar is modified, for instance via the inclusion of strong localised interactions. This leaves the door open for the construction of {\it lemon aGUT} models, based on the inclusion of ad-hoc non-minimal terms. Finally, some models may benefit from the addition of supersymmetry, which ensures the vanishing of the gauge-scalar potential. Hence, we identified a second $\SU(8)$ model, where supersymmetry allows us to render the orbifold stable and the scalar potential is generated by a supersymmetry breaking mechanism.

\section*{Acknowledgements}
ASC is supported in part by the National Research Foundation (NRF) of South Africa. GC thanks the LPTHE, Sorbonne University, for their hospitality while this work was completed.

\appendix 

\section*{Appendices}

%%%%%%%%%%%%%%%%%%%%%%%%%%%%%%%%%%%%%%%%%%%%%
\section{Parity definition on orbifolds}
\label{ParityDefinition}
%%%%%%%%%%%%%%%%%%%%%%%%%%%%%%%%%%%%%%%%%%%%%

\paragraph{}
In this Appendix, we will derive the general form of the orbifold parities depending on the gauge group we consider in the bulk. The parity matrices are elements of the Cartan subalgebra of the gauge group with an overall phase
\begin{equation}
    P = e^{i \theta_0}\ \Omega(\theta_j)\,,
\end{equation}
to which the additional condition $P \cdot P=\mathbb{1}$ is imposed. 

%%%%%%%%%%%%%%%%%%%%%%%%%%%%%%%%
\subsection{$\SU(N)$}
%%%%%%%%%%%%%%%%%%%%%%%%%%%%%%%%

The Cartan subalgebra is generated by the diagonal generators
\begin{equation}
    X_j^C = \mbox{diag} (0, \dots 0, 1, 0, \dots -1)\,, \;\; j = 1, \dots N-1\,,
\end{equation}
where the ``1'' is in the $j$-th position. Hence,
\begin{equation}
    \Omega_{\SU(N)} = \mbox{diag} (e^{i \theta_1},\ \dots , e^{i \theta_{N-1}}, \prod_j e^{-i \theta_i})\,.
\end{equation}
This implies that the most general parity matrix will be diagonal, and can only have $\pm 1$ entries. To obtain the most general combination, one can choose $\theta_j = - \theta_0$ or $\pi-\theta_0$, so that
\begin{equation}
    P\left(\theta_0,\theta_i \in \{-\theta_0,\pi-\theta_0 \} \right) ={\rm diag}(\underbrace{-1,\cdots,-1}_{\alpha},\underbrace{+1,\cdots,+1}_{N-1-\alpha}, (-1)^{\alpha} e^{i N \theta_0})\,,\quad \alpha \in [\![  1,N-1 ]\!]
\end{equation}
where we can choose $N \theta_0 = \pi$ or $0$. In summary, the most general $\SU(N)$ parity matrix reads
\begin{equation}
    P_{\SU(N)} = {\rm diag}(\underbrace{1,\cdots,1}_{A},\underbrace{-1,\cdots,-1}_{N-A}) \,,
\end{equation}
with $A \in [\![  1,N ]\!]$. This parity yields the breaking pattern
\begin{equation}
    \SU(N) \to \SU(A) \times \SU(N-A) \times \U(1)\,.
\end{equation}

%%%%%%%%%%%%%%%%%%%%%%%%%%%%%%%%%%%%%
\subsection{$\Sp(2N)$}
%%%%%%%%%%%%%%%%%%%%%%%%%%%%%%%%%%%%%

The generators of the Cartan subalgebra are given by:
\begin{equation}
    X_j = \mbox{diag} (0, \dots 0, 1, 0, \dots 0) \otimes \begin{pmatrix} 1 & 0 \\ 0 & -1 \end{pmatrix}\,, \;\; j = 1, \dots N\,,
\end{equation}
where the ``1'' is in the j-th position. Hence, a general element of the Cartan subalgebra reads
\begin{equation}
    \Omega_{\Sp(2N)} = \mbox{diag} (e^{i \theta_1}\,, \dots e^{i \theta_N}, e^{-i \theta_1}, \dots e^{-i \theta_N}) \,.
\end{equation}
There are two independent choices for the definition of a parity:
\begin{enumerate}
        \item[I)] One can choose $\theta_0=0$ and $\theta_i = 0$ or $\pi$, thus obtaining:
        \begin{equation}
        P^{\rm I}_{\Sp(2N)} = {\rm diag}(\underbrace{-1,\cdots,-1}_{A},\underbrace{+1,\cdots,+1}_{N-A})\otimes \begin{pmatrix} 1 & 0 \\ 0 & 1 \end{pmatrix}=P_{\SU(N)}\otimes \begin{pmatrix} 1 & 0 \\ 0 & 1 \end{pmatrix}\,,
        \end{equation}
        which corresponds to the breaking
        \begin{equation}
            \Sp(2N) \to \Sp(2A) \times \Sp(2(N-A)) \,.
        \end{equation}
        \item[II)] One can choose $\theta_0=\pi/2$ and $\theta_i = \pm \pi/2$, thus obtaining
        \begin{equation}
        P^{\rm II}_{\Sp(2N)} = {\rm diag}(\underbrace{-1,\cdots,-1}_{A},\underbrace{+1,\cdots,+1}_{N-A})\otimes \begin{pmatrix} 1 & 0 \\ 0 & -1 \end{pmatrix}=P_{\SU(N)}\otimes \begin{pmatrix} 1 & 0 \\ 0 & -1 \end{pmatrix}\,,
        \end{equation}
        yielding the pattern
        \begin{equation}
            \Sp(2N) \to \SU(N) \times \U(1) \,.
        \end{equation}
        The fact that the breaking does not depend on $A$ means that one can always flip signs in the parity, leading to:
        \begin{equation} \label{eqapp:PIISp2N}
        P^{\rm II}_{\Sp(2N)} = {\rm diag}(\underbrace{1,\cdots,1}_{N})\otimes \begin{pmatrix} 1 & 0 \\ 0 & -1 \end{pmatrix}\,.
        \end{equation}
\end{enumerate}
Hence, the $\Sp(2N)$ case features two qualitatively different parity matrices.

%%%%%%%%%%%%%%%%%%%%%%%%%%%%%%
\subsection{$\SO(2N)$}
%%%%%%%%%%%%%%%%%%%%%%%%%%%%%%

The Cartan subalgebra is generated by the following $N$ anti-symmetric matrices:
\begin{equation}
    X_j = \sigma_2 \otimes \mbox{diag} (0, \dots 0, 1, 0 \dots 0)\,,  \;\; j = 1, \dots N\,,
\end{equation}
where, again, the ``1'' is in the j-th position. It is more convenient to rewrite it in a form resembling the structure for $\Sp(2N)$, so that
\begin{equation}
    \tilde{X}_j =  \mbox{diag} (0, \dots 0, 1, 0 \dots 0) \otimes \sigma_2\,,  \;\; j = 1, \dots N\,.
\end{equation}
The generators can be written in terms of four $N\times N$ blocks as
\begin{equation}
    X = \begin{pmatrix} A & B \\ -B^T & D \end{pmatrix} \,,
\end{equation}
where $A^T= -A$ and $D^T = -D$. We can construct the $\SO(2N)$ transformation,
\begin{equation}
    \Omega_{\SO(2N)} = \sum_i \mbox{diag} (0, \dots 0, 1, 0 \dots 0) \otimes e^{i \theta_j \sigma_2}\,,
\end{equation}
where
\begin{equation}
    e^{i \theta_i \sigma_2} = \begin{pmatrix} \cos \theta_i & \sin \theta_i \\ - \sin \theta_i & \cos \theta_i \end{pmatrix}\,.
\end{equation}
We notice that $\Omega_{\SO(2N)} \cdot \Omega_{\SO(2N)}$ is made of blocks proportional to $e^{i 2 \theta_j \sigma_2}$. There are two choices of $\theta_j$ to make it proportional to the identity:
\begin{equation}
    e^{i 2 \theta_j \sigma_2} = - \mathbb{1}_2\;\; \mbox{for} \;\; \theta_j = \frac{\pi}{2},\ \frac{3\pi}{2}\,, \qquad  e^{i 2 \theta_j \sigma_2} = \mathbb{1}_2\;\; \mbox{for} \;\; \theta_j = 0,\ \pi\,.
\end{equation}
Hence, there are two ways to construct a parity matrix:
\begin{enumerate}
        \item[I)] By choosing $\theta_j = 0$ or $\pi$, we can build diagonal matrices as follows:
        \begin{equation}
            P_{\SO(2N)}^{\rm I} = \Omega_{\SO(2N)} = P_{\SU(N)} \otimes \begin{pmatrix} 1 & 0 \\ 0 & 1 \end{pmatrix}  \,,
        \end{equation}
        which gives rise to the breaking
        \begin{equation}
            \SO(2N) \to \SO(2A) \times \SO(2(N-A))\,.
        \end{equation}

        \item[II)] By choosing $\theta_j = \pi/2$ or $3\pi/2$, we define
        \begin{equation}
            P_{\SO(2N)}^{\rm II} = i \Omega_{\SO(2N)} = P_{\SU(N)} \otimes \begin{pmatrix} 0 & -i \\ i & 0 \end{pmatrix} \,,
            \label{eq:PII_SO(2N)app}
        \end{equation}
        which triggers the breaking
        \begin{equation}
            \SO(2N) \to \SU(N)\times \U(1)\,.
        \end{equation}
         Note, however, that this parity will act non-trivially on the four blocks of the generators. In fact, we can write explicitly
        \begin{equation}
             P^{\rm II}_{\SO(2N)} = \begin{pmatrix}
            0 & - i P_{\SU(N)} \\ i P_{\SU(N)} & 0
        \end{pmatrix}\,.
        \end{equation}
        Hence,
        \begin{equation}
            P^{\rm II}_{\SO(2N)} \cdot \begin{pmatrix}
            A & B \\ -B^T & D
        \end{pmatrix} \cdot P^{\rm II}_{\SO(2N)} = \begin{pmatrix}
            P_{\SU(N)}\cdot D \cdot P_{\SU(N)} & - P_{\SU(N)}\cdot B^T \cdot P_{\SU(N)}  \\
            P_{\SU(N)}\cdot B \cdot P_{\SU(N)} & P_{\SU(N)}\cdot A \cdot P_{\SU(N)}
        \end{pmatrix}\,. 
        \end{equation}
        One can define combinations of the matrices $A$, $B$ and $D$ with definite parities as follows:
        \begin{equation}
            A_\pm = A \pm D\,, \qquad B = B_a + B_s\,,
        \end{equation}
        where the subscripts indicate the symmetric and antisymmetric components of $B$ (i.e., $B_s^T = B_s$ and $B_a^T = -B_a$), such that
        \begin{eqnarray}
            A_\pm &\to& \pm P_{\SU(N)} \cdot A_\pm \cdot P_{\SU(N)}\,, \\
            B_{a/s} &\to& \pm P_{\SU(N)} \cdot B_{a/s} \cdot P_{\SU(N)} \,.
        \end{eqnarray}
        In cases where this type of parity is employed, it is better to use $A_\pm$ and $B_{a/s}$ as defining blocks for the generators. Note also that this is compatible with the other type of parity, for which all blocks will receive the same parity assignments. Thus, we will use the notations
        \begin{equation}
            X=\underbrace{\begin{pmatrix}
            A_{+}/2&B_{s}\\
            -B_{s}&A_{+}/2
           \end{pmatrix}}_{X_{+}}
            +
           \underbrace{\begin{pmatrix}
             A_{-}/2&B_{a}\\
            B_{a}&-A_{-}/2   
            \end{pmatrix}}_{X_{-}},
        \end{equation}
        where $X_{\pm}$ schematically transform under $P^{\rm II}$ as
        \begin{eqnarray}
            X_{+} &\to&  P_{\SU(N)}\cdot X_{+} \cdot P_{\SU(N)}\,, \\
            X_{-} &\to& -P_{\SU(N)}\cdot X_{-} \cdot P_{\SU(N)}\,.
        \end{eqnarray}
        The projection matrices corresponding to the action of $P^{\rm II}$ on $X_{\pm}$ are
        \begin{equation}
         \begin{pmatrix}
            1 &1\\
            1&1 
        \end{pmatrix}\:\:\:
        \text{and}\:\:\:
         \begin{pmatrix}
            -1 &-1\\
            -1&-1 
        \end{pmatrix} \,.
        \end{equation}
        Also, the adjoint Adj will decompose as
        \begin{equation}
          \mbox{Adj} \to \mbox{Adj} \oplus S_x \oplus \bar{S}_{-x}\,.
        \end{equation}
\end{enumerate}

%%%%%%%%%%%%%%%%%%%%%%%%%%%%%%%
\subsection{$\SO(2N+1)$}
%%%%%%%%%%%%%%%%%%%%%%%%%%%%%%%

For odd-dimension orthogonal groups, the Cartan subalgebra has the same dimension as the corresponding even $\SO(2N)$ group. Hence, the generators have the same form, with an additional entry that remains empty. Because of the odd dimension, one can only define parities of type $P^{\rm I}_{\SO}$ similar to the even case.

%%%%%%%%%%%%%%%%%%%%%%%%%%%%%%%%%%%%%%%%%%%%%%%%%%%%%%%%
\section{General formulas for the effective potential}
\label{Potential}
%%%%%%%%%%%%%%%%%%%%%%%%%%%%%%%%%%%%%%%%%%%%%%%%%%%%%%%%

In this Appendix, we will detail the computation of the Coleman-Weinberg effective potential for each gauge group in the bulk.

%%%%%%%%%%%%%%%%%%%%%%%%%%
\subsection{$\SU(N)$}
%%%%%%%%%%%%%%%%%%%%%%%%%%

We consider here the most general parity assignment, leading to a four-block symmetry breaking pattern \cite{Haba_2004}. From Eq. \eqref{paritiesSUN}, we can deduce that the gauge-scalar zero modes always transform as bi-fundamental representations. Without loss of generality, their VEVs can be expressed in diagonal block form, with dimensions $n_{ps} = \mbox{min} (p,s)$ and $n_{qr} = \mbox{min} (q,r)$, respectively. As a starter, we can consider a matrix restricted to only rows and columns where a VEV appears
\begin{equation}
   (P_1,P_2) (A_{\mu}^*) = \begin{pNiceMatrix}[first-row,last-col=5,code-for-first-row=\scriptstyle,code-for-last-col=\scriptstyle] 
     n_{ps} & n_{qr} & n_{qr} & n_{ps} &  \\
      (+,+) &  (+,-) & (-,+) & (-,-) & n_{ps}  \\
     (+,-) & (+,+) & (-,-) & (-,+) & n_{qr}\\
     (-,+) & (-,-) & (+,+) & (+,-) & n_{qr}  \\
     (-,-) & (-,+) & (+,-) & (+,+) & n_{ps}  \\
    \end{pNiceMatrix}\,.
\end{equation}
The VEVs can be written in terms of a `charge' matrix
\begin{equation}
    Q =  \begin{pNiceMatrix}[first-row,last-col=5,code-for-first-row=\scriptstyle,code-for-last-col=\scriptstyle] n_{ps} &  n_{qr} &  n_{qr}& n_{ps} & \\  
    0 &  0 & 0 & \mbox{diag}[a_i] & n_{ps}\\
    0 & 0 & \mbox{diag}[b_j] & 0 & n_{qr} \\
    0 & \mbox{diag}[b_j] & 0 & 0 & n_{qr}\\
    \mbox{diag}[a_i] & 0 & 0 & 0 & n_{ps}\\
    \end{pNiceMatrix}\,.
\end{equation}
This block matrix can be diagonalised via
\begin{equation}
    V = \frac{1}{\sqrt{2}} \begin{pmatrix}
    \mathbb{1}_{n_{ps}} & 0 & 0 & -\mathbb{1}_{n_{ps}} \\ 
    0 & \mathbb{1}_{n_{qr}} & -\mathbb{1}_{n_{qr}} & 0 \\ 
    0 & \mathbb{1}_{n_{qr}} & \mathbb{1}_{n_{qr}} & 0 \\
    \mathbb{1}_{n_{ps}} & 0 & 0 & \mathbb{1}_{n_{ps}}
    \end{pmatrix}\,,    
\end{equation}
so that
\begin{equation}
    Q^{\mbox{diag}} = V \cdot Q \cdot V^T = \begin{pmatrix}
    -\mbox{diag}[a_i] & 0 & 0 & 0 \\
    0 & -\mbox{diag}[b_j] & 0 & 0 \\
    0 & 0 & \mbox{diag}[b_j] & 0 \\
    0 & 0 & 0 & \mbox{diag}[a_i]
    \end{pmatrix}\,.
\end{equation}
Note that, once applied to $A_\mu$, the rotation $V$ only mixes states with the same KK mass, as $(+,+)$ and $(-,-)$ have KK mass $n/R$, while $(+,-)$ and $(-,+)$ have KK mass $(n+1/2)/R$. In the diagonal basis, the KK spectrum coming from the various blocks can be derived by computing the VEV `charge' of the various components:
\begin{equation}
   [Q, A_\mu^\ast] \to \begin{pNiceMatrix}[first-row,last-col=5,code-for-first-row=\scriptstyle,code-for-last-col=\scriptstyle] n_{ps} & n_{qr}&  n_{qr} & n_{ps}& \\  
    n - a_i + a_k &  n+\frac{1}{2} -a_i+b_l & n+\frac{1}{2} - a_i - b_l & n-a_i-a_k & n_{ps}\\
    n+\frac{1}{2}-b_j+a_k & n-b_j+b_l & n-b_j-b_l & n+\frac{1}{2} -b_j-a_k & n_{qr}\\
    n+\frac{1}{2}+b_j+a_k & n+b_j+b_l & n+b_j-b_l & n+\frac{1}{2} + b_j - a_k & n_{qr}\\
    n+a_i+a_k & n+\frac{1}{2} +a_i+b_l & n+\frac{1}{2} + a_i-b_l & n+a_i-a_k & n_{ps}\\
    \end{pNiceMatrix}\,.
\end{equation}
The components we discarded also receive a mass contribution from the VEV `charges'. If we consider the first row of the matrix, the $(+,+)$ block has $|p-s|$ off-diagonal components that receive charges $a_i$, while the $(+,-)$ block has $|q-r|$ components receiving charges $a_i$. Similarly, in the second row, the $(+,+)$ block has $|q-r|$ components receiving charges $b_j$ and the $(-,+)$ block has $|p-s|$ components receiving charges $b_j$.

Putting all the results together, the most general potential from the adjoint representation reads
\begin{align}
     \left. \mathcal{V}_{Adj}\right|_{\SU(N)} =&  \sum_{i,k=1}^{n_{ps}} \left( \mathcal{F}^+ (a_i+a_k) + \mathcal{F}^+ (a_i-a_k) \right) +  \sum_{j,l=1}^{n_{qr}} \left( \mathcal{F}^+ (b_j+b_l) + \mathcal{F}^+ (b_j-b_l) \right) \nonumber \\
    & + 2 \sum_{i=1}^{n_{ps}} \sum_{j=1}^{n_{qr}} \left( \mathcal{F}^- (a_i+b_j) +  \mathcal{F}^- (a_i-b_j) \right) \nonumber \\
    & + 2 |p-s| \left( \sum_{i=1}^{n_{ps}} \mathcal{F}^+ (a_i) + \sum_{j=1}^{n_{qr}} \mathcal{F}^- (b_j) \right) + 2 |q-r| \left( \sum_{i=1}^{n_{ps}} \mathcal{F}^- (a_i) + \sum_{j=1}^{n_{qr}} \mathcal{F}^+ (b_j) \right)\,,\label{eq:VeffgenSUN}
\end{align}
in agreement with Ref.~\cite{Haba_2004}. We will use the same VEV `charge' technique for the other general cases.

%%%%%%%%%%%%%%%%%%%%%%%%%%%%%%%%%
\subsection{$\Sp(2N)$}
%%%%%%%%%%%%%%%%%%%%%%%%%%%%%%%%%

We need to distinguish three cases, depending on the type of parity applied on the two boundaries: I+I, I+II, II+II.

%%%%%%%%%%%%%%%%%%%%%%%%%%%%%%%%%
\subsubsection{Case I+I}
%%%%%%%%%%%%%%%%%%%%%%%%%%%%%%%%%

In this case, the most general configuration breaks $\Sp(2N)$ into four subgroups. The most general parities for the gauge multiplet read
\begin{equation}
      (P_1,P_2)(A_{\mu})= \left( \begin{array}{c|c}
        \begin{array}{cccc} {(+,+)} &  (+,-) & (-,+) & {(-,-)} \\
    (+,-) & {(+,+)} & {(-,-)} & (-,+) \\
    (-,+) & {(-,-)} & {(+,+)} & (+,-) \\
    {(-,-)} & (-,+) & (+,-) & {(+,+)} \\\end{array}& 
    \begin{array}{cccc} {(+,+)} &  (+,-) & (-,+) & {(-,-)} \\
    (+,-) & {(+,+)} & {(-,-)} & (-,+) \\
    (-,+) & {(-,-)} & {(+,+)} & (+,-) \\
    {(-,-)} & (-,+) & (+,-) & {(+,+)} \\\end{array}\\ 
    \hline \begin{array}{cccc} {(+,+)} &  (+,-) & (-,+) & {(-,-)} \\
    (+,-) & {(+,+)} & {(-,-)} & (-,+) \\
    (-,+) & {(-,-)} & {(+,+)} & (+,-) \\
    {(-,-)} & (-,+) & (+,-) & {(+,+)} \\\end{array} & 
    \begin{array}{cccc} {(+,+)} &  (+,-) & (-,+) & {(-,-)} \\
    (+,-) & {(+,+)} & {(-,-)} & (-,+) \\
    (-,+) & {(-,-)} & {(+,+)} & (+,-) \\
    {(-,-)} & (-,+) & (+,-) & {(+,+)} \\\end{array} \end{array}\right)\,. 
\end{equation}
The gauge-scalars transform as bi-fundamental representations of $\Sp(2p) \times \Sp(2s)$ and of $\Sp(2q)\times \Sp(2r)$. We recall that the structure of the generators is as follows
\begin{equation}
   X_{\Sp(2N)}= \left( \begin{array}{c|c}
    A & B \\ \hline
    C & -A^T \end{array} \right)\,,
\end{equation}
where $B$ and $C$ are symmetric matrices.
Each gauge-scalar should allow for $n_{ps}$ and $n_{qr}$ independent VEVs
\begin{equation} \label{eq:QSp2NII}
       Q = i \left( \begin{array}{cccc|cccc}
     0 &  0 & 0 & {\mbox{diag}[a_i]} & 0 & 0 & 0 & 0 \\
    0 & 0 & {\mbox{diag}[b_j]} & 0 & 0 & 0 & 0 & 0\\
    0 &{-\mbox{diag}[b_j]} & 0 & 0 & 0 & 0 & 0 & 0\\
    {-\mbox{diag}[a_i]} & 0 & 0 & 0 & 0 & 0 & 0 & 0\\ \hline
    0 & 0 & 0 & 0 & 0 &  0 & 0 & {\mbox{diag}[a_i]} \\
    0 & 0 & 0 & 0 &0 & 0 & {\mbox{diag}[b_j]} & 0 \\
    0 & 0 & 0 & 0 &0 & {-\mbox{diag}[b_j]} & 0 & 0 \\
    0 & 0 & 0 & 0 &{-\mbox{diag}[a_i]} & 0 & 0 & 0 \\\end{array}\right)\,.
\end{equation}
The complex VEV is chosen for convenience, and we can check that $a_i = a$ and $b_j = b$ preserves a $\Sp(2n_{ps}) \times \Sp(2 n_{qr})$ subgroup, hence, it is the most general maximally breaking VEV structure. The above charge can be diagonalised via a block matrix:
\begin{equation}
    V = \left( \begin{array}{c|c} \mathcal{V} & 0 \\ \hline 0 & \mathcal{V} \end{array} \right) \qquad \mbox{with} \quad \mathcal{V} = \begin{pmatrix}
    \mathbb{1}_{n_{ps}} & 0 & 0 & -i \mathbb{1}_{n_{ps}} \\
    0 & \mathbb{1}_{n_{qr}} & -i \mathbb{1}_{n_{qr}} & 0 \\
    0 & -i \mathbb{1}_{n_{qr}} & \mathbb{1}_{n_{qr}} & 0 \\
    -i \mathbb{1}_{n_{ps}} & 0 & 0 & \mathbb{1}_{n_{ps}}
    \end{pmatrix}\,.
\end{equation}
The diagonal VEV `charge' matrix reads:
\begin{equation}
\label{eq:Q diagonal}
   V \cdot Q \cdot V^\dagger = \left( \begin{array}{cccc|cccc}
    {- \mbox{diag}[a_i]} &0&0&0&0&0&0&0\\
    0 &  {- \mbox{diag}[b_j]} & 0 & 0 &0&0&0&0\\
    0 & 0 & {\mbox{diag}[b_j]} & 0  &0&0&0&0\\
    0 & 0 & 0 & { \mbox{diag}[a_i]} &0&0&0&0 \\  \hline 0&0&0&0& {- \mbox{diag}[a_i]} &0&0&0\\
    0&0&0&0& 0 &  {- \mbox{diag}[b_j]} & 0 & 0 \\
    0&0&0&0& 0 & 0 & {\mbox{diag}[b_j]} & 0  \\
    0&0&0&0& 0 & 0 & 0 &{ \mbox{diag}[a_i]}  \\\end{array}\right)\,.
\end{equation}
The structure of the VEV `charges' is the same as for the $\SU(N)$ case. We can then conclude that block $A$ provides the same contribution to the potential as the adjoint of $\SU(N)$ and each $B$ and $C$ blocks give half the contribution of the adjoint of $\SU(N)$. This can be deduced by the fact that diagonal elements have zero charge, so only the off-diagonal ones contribute. Thus the symmetric part of the $\SU(N)$ adjoint constitute half of the total effective potential. Summing the contribution of $A$, $B$ and $C$, we obtain
\begin{equation}
    \left. \mathcal{V}_{Adj}^{\rm I+I}\right|_{\Sp(2N)} = 2\  \left.\mathcal{V}_{Adj}\right|_{\SU(N)}\,.
\end{equation}

%%%%%%%%%%%%%%%%%%%%%%%%%%%%%%%%%
\subsubsection{Case I+II}
%%%%%%%%%%%%%%%%%%%%%%%%%%%%%%%%%

In this case, two $\U(n)$ symmetries are preserved. The most general parity configuration reads
        \begin{equation}
        (P_1,P_2)(A_\mu) = \left( \begin{array}{c|c} \begin{array}{cc} {(+,+)} & (+,-) \\ (+,-) & {(+,+)} \end{array} & \begin{array}{cc} (-,+) & {(-,-)} \\ {(-,-)} & (-,+) \end{array} \\ \hline \begin{array}{cc} (-,+) & {(-,-)} \\ {(-,-)} & (-,+) \end{array} & \begin{array}{cc} {(+,+)} & (+,-) \\ (+,-) & {(+,+)} \end{array} \end{array}\right)\,, 
    \end{equation}
hence, the most general VEV structure is as follows
\begin{equation}
    Q = \left( \begin{array}{cc|cc} 
    0 & 0 & 0 & \mbox{diag}[a_i]\\
    0 & 0 & \mbox{diag}[a_i]  & 0 \\\hline 0 & \mbox{diag}[a_i] & 0 & 0\\
    \mbox{diag}[a_i]  & 0 & 0 & 0\\
    \end{array}\right)\,.
\end{equation}
This VEV `charge' matrix can be diagonalised to
\begin{equation}
    Q^{\rm diag} = V\cdot Q \cdot V^T = \left( \begin{array}{cc|cc} 
    -\mbox{diag}[a_i] & 0 & 0 & 0 \\
    0 & \mbox{diag}[a_i]  & 0 & 0  \\\hline  0& 0&   -\mbox{diag}[a_i] & 0\\
    0& 0&  0 & \mbox{diag}[a_i]\\
    \end{array}\right)\,.
\end{equation}
We can immediately infer that the $A$ block of the $\Sp(2N)$ adjoint will contribute as the adjoint of $\SU(N)$ with only one type of VEV. The $B$ and $C$ blocks together contribute as an $\SU(N)$ adjoint with one parity flipped. Concretely, the effective potential is the $\SU(N)$ one with the exchange of  $\mathcal{F}^+$ and  $ \mathcal{F}^-$. All in all, we find:
\begin{align}
    \left.\mathcal{V}^{\rm I+II}_{Adj}\right|_{\Sp(2N)} = & \frac{1}{16} \left[\sum_{m,n=1}^{n_{pq}} \left( \mathcal{F}^+ (2a_m + 2a_n)  + \mathcal{F}^+ (2a_m - 2a_n) \right) + 2 |p-q| \; \sum_{m=1}^{n_{pq}}\mathcal{F}^+ (2a_m) \right]\,,
\end{align}
where $n_{pq} = \mbox{min}[p,q]$.

%%%%%%%%%%%%%%%%%%%%%%%%%%%%%%%%%
\subsubsection{Case II+II}
%%%%%%%%%%%%%%%%%%%%%%%%%%%%%%%%%

We have seen that the most general configurations for the parities lead to:
\begin{equation}
    \label{eq:A_II+II}
        (P_1,P_2)(A_\mu) = \left( \begin{array}{c|c} \begin{array}{cc} {(+,+)} & (+,-) \\ (+,-) & {(+,+)} \end{array} & \begin{array}{cc} {(-,-)} & (-,+) \\ (-,+) & {(-,-)} \end{array} \\ \hline \begin{array}{cc} {(-,-)} & (-,+) \\ (-,+) &{(-,-)} \end{array} & \begin{array}{cc} {(+,+)} & (+,-) \\ (+,-) & {(+,+)} \end{array} \end{array}\right)\,,
    \end{equation}
such that the gauge-scalars $\varphi_p$ and $\varphi_q$ can be found in the symmetric representations of $\U(p)$ and $\U(q)$, respectively. Without loss of generality, we can define the VEVs of the gauge-scalars in the following way:
\begin{equation}
    \langle \varphi_p \rangle = {\rm diag}[a_i] \,, ~~~~~ \langle \varphi_q \rangle = {\rm diag}[b_j]\,.
\end{equation}
Thus the charge matrix reads
\begin{equation}
    Q = \left( \begin{array}{cc|cc} 
    0 & 0&  \mbox{diag}[a_i] & 0\\
    0 & 0&  0  & \mbox{diag}[b_j]  \\\hline  \mbox{diag}[a_i] & 0 &0&0\\
    0  & \mbox{diag}[b_j]  &0&0 \\
    \end{array}\right)\,.
\end{equation}
This VEV `charge' matrix can also be diagonalised to simplify the computation of the contributions to the effective potential. We end up with a very similar result as in $\SU(N)$, except that the off-diagonal terms are not there as we are working with symmetric representations. The effective potential is then
\begin{equation}
\begin{split}
     \left.\mathcal{V}^{{\rm II+II}}_{Adj}\right|_{\Sp(2N)} = &  \sum_{i,k=1}^{p} \left( \mathcal{F}^+ (a_i+a_k) + \mathcal{F}^+ (a_i-a_k) \right) +  \sum_{j,l=1}^{q} \left( \mathcal{F}^+ (b_j+b_l) + \mathcal{F}^+ (b_j-b_l) \right)  \\
    & + 2 \sum_{i=1}^{p} \sum_{j=1}^{q} \left( \mathcal{F}^- (a_i+b_j) +  \mathcal{F}^- (a_i-b_j) \right) \,.
\end{split}
\end{equation}

%%%%%%%%%%%%%%%%%%%%%%%%%%%%%%%%%
\subsection{$\SO(2N)$}
%%%%%%%%%%%%%%%%%%%%%%%%%%%%%%%%%

As already described for the $\Sp(2N)$ scenario, there are three independent cases based on the parity types.

%%%%%%%%%%%%%%%%%%%%%%%%%%%%%%%%%
\subsubsection{Case I+I}
%%%%%%%%%%%%%%%%%%%%%%%%%%%%%%%%%

The parity assignments of the adjoint representation are the same as in the $\Sp(2N)$ case. We note that the gauge-scalars transform as bi-fundamentals of $\SO(2p) \times \SO(2s)$  and of $\SO(2q)\times \SO(2r)$ and we can choose the VEV in the off-diagonal blocks, such that
\begin{equation} 
       Q = i \left( \begin{array}{cccc|cccc}
    0&0&0&0& 
    0 &  0 & 0 & {\mbox{diag}[a_i]} \\
    0&0&0&0&0 & 0 & {\mbox{diag}[b_j]} & 0 \\
    0&0&0&0&0 & {\mbox{diag}[b_j]} & 0 & 0 \\
    0&0&0&0&{\mbox{diag}[a_i]} & 0 & 0 & 0 \\ 
    \hline 
   0 &  0 & 0 & {-\mbox{diag}[a_i]} &0&0&0&0 \\
    0 & 0 & {-\mbox{diag}[b_j]} & 0  &0&0&0&0\\
    0 & {-\mbox{diag}[b_j]} & 0 & 0  &0&0&0&0\\
    {-\mbox{diag}[a_i]} & 0 & 0 & 0  &0&0&0&0\\\end{array} \right)\,. 
\end{equation}
To find the potential, we first need to diagonalise the VEV `charge' via
\begin{equation}
    V = \frac{1}{\sqrt{2}}\left(
\begin{array}{cccccccc}
 \mathbb{1}_{n_{ps}} & 0 & 0 & 0 & 0 & 0 & 0 & -i \mathbb{1}_{n_{ps}} \\
 0 & \mathbb{1}_{n_{qr}} & 0 & 0 & 0 & 0 & -i \mathbb{1}_{n_{qr}} & 0 \\
 0 & 0 & \mathbb{1}_{n_{qr}} & 0 & 0 & -i \mathbb{1}_{n_{qr}} & 0 & 0 \\
 0 & 0 & 0 & \mathbb{1}_{n_{qr}} & -i \mathbb{1}_{n_{ps}} & 0 & 0 & 0 \\
 0 & 0 & 0 & -i \mathbb{1}_{n_{ps}} & \mathbb{1}_{n_{ps}} & 0 & 0 & 0 \\
 0 & 0 & -i \mathbb{1}_{n_{qr}}& 0 & 0 & \mathbb{1}_{n_{qr}} & 0 & 0 \\
 0 & -i \mathbb{1}_{n_{qr}}& 0 & 0 & 0 & 0 & \mathbb{1}_{n_{qr}} & 0 \\
 -i \mathbb{1}_{n_{ps}}& 0 & 0 & 0 & 0 & 0 & 0 & \mathbb{1}_{n_{ps}} \\
\end{array}
\right)\, ,
\end{equation}
which gives the diagonal matrix
\begin{equation}
    Q^{\rm diag}=\left(
\begin{array}{cccc|cccc}
 {\ -\mbox{diag}[a_i]} & 0 & 0 & 0 & 0 & 0 & 0 & 0 \\
 0 & {-\ \mbox{diag}[b_j]} & 0 & 0 & 0 & 0 & 0 & 0 \\
 0 & 0 & {-\ \mbox{diag}[b_j]} & 0 & 0 & 0 & 0 & 0 \\
 0 & 0 & 0 & {-\ \mbox{diag}[a_i]} & 0 & 0 & 0 & 0 \\ \hline
 0 & 0 & 0 & 0 & {\ \mbox{diag}[a_i]} & 0 & 0 & 0 \\
 0 & 0 & 0 & 0 & 0 &{\ \mbox{diag}[b_j]} & 0 & 0 \\
 0 & 0 & 0 & 0 & 0 & 0 & {\ \mbox{diag}[b_j]} & 0 \\
 0 & 0 & 0 & 0 & 0 & 0 & 0 & {\ \mbox{diag}[a_i]} \\
\end{array}
\right) \,.
\end{equation}
The potential is then computed in a similar way as in the $\Sp(2N)$ case. The off-diagonal block gives the same contribution as one $\SU(N)$ adjoint, while the diagonal blocks give a half of an $\SU(N)$ contribution due to their antisymmetric nature. The final result becomes
\begin{equation}
    \left.\mathcal{V}^{{\rm I+I}}_{Adj}\right|_{\SO(2N)} = 2\  \left.\mathcal{V}_{Adj}\right|_{\SU(N)}\,.
\end{equation}

%%%%%%%%%%%%%%%%%%%%%%%%%%%%%%%%%
\subsubsection{Case I+II}
\label{subsec: SO(2N) I+II}
%%%%%%%%%%%%%%%%%%%%%%%%%%%%%%%%%

\paragraph{}
We will consider $P_1$ to be the type II parity and $P_2$ to be of the type I. Defining $A_{\mu}^+$ and $A_{\mu}^-$ as the components of the adjoint associated to $X_+$ and $X_-$, we find the following parity assignments:
\begin{equation}
    (P_1,P_2)(A^{+}_\mu) =\left( \begin{array}{c|c} \begin{array}{cc} {(+,+)} & (+,-) \\ (+,-) & {(+,+)} \end{array} & \begin{array}{cc} {(+,+)} & (+,-) \\ (+,-) & {(+,+)} \end{array} \\ \hline \begin{array}{cc} {(+,+)} & (+,-) \\ (+,-) & {(+,+)} \end{array} & \begin{array}{cc} {(+,+)} & (+,-) \\ (+,-) & {(+,+)} \end{array} \end{array}\right)\,, 
\end{equation}  
\begin{equation}
    (P_1,P_2)(A^{-}_\mu) = \left( \begin{array}{c|c} \begin{array}{cc} {(-,+)} & (-,-) \\ (-,-) & {(-,+)} \end{array} & \begin{array}{cc} {(-,+)} & (-,-) \\ (-,-) & {(-,+)} \end{array} \\ \hline \begin{array}{cc} {(-,+)} & (-,-) \\ (-,-) & {(-,+)} \end{array} & \begin{array}{cc} {(-,+)} & (-,-) \\ (-,-) & {(-,+)} \end{array} \end{array}\right) \,.
            \end{equation}
We notice here that only $A_{\mu}^-$ contains gauge-scalar zero modes, living in the bi-fundamental representation of $\U(p)\times \U(q)$. Hence, the most general structure for the VEV of $A_5$ is
\begin{equation}
    Q = i \left( \begin{array}{cc|cc} 
    0&0& 0 & \mbox{diag}[a_i]\\
    0&0&-\mbox{diag}[a_i]  & 0 \\\hline 0 & \mbox{diag}[a_i] &0&0\\
    -\mbox{diag}[a_i]  & 0 &0&0 \\
    \end{array}\right)\,.
\end{equation}
Once again we can diagonalise the VEV `charge', resulting in
\begin{equation}
    Q^{\rm diag}= \left( \begin{array}{cc|cc} 
     -\mbox{diag}[a_i] & 0 &0&0\\
    0 & -\mbox{diag}[a_i] &0&0\\\hline 0&0&  \mbox{diag}[a_i] & 0\\
    0&0& 0 & \mbox{diag}[a_i]   \\
    \end{array}\right)\,.
\end{equation}
The generated spectrum for $A^{\pm}_{\mu}$ will then be computed using $[Q^{\rm diag}, A^{\pm}_{\mu}]$
\begin{equation}
\label{eq:so(2n)A+}
    [Q^{\rm diag},A^{+}_\mu] = \left( \begin{array}{c|c} \begin{array}{cc} 0 &  n+\frac{1}{2} -a_i+a_j \\  n+\frac{1}{2}-a_i+a_j & 0 \end{array} & \begin{array}{cc}n- a_i - a_j & n+\frac{1}{2}-a_i-a_j \\  n+\frac{1}{2}-a_i-a_j & n -a_i-a_j \end{array} \\ \hline \begin{array}{cc}  n+a_i+a_j & n+\frac{1}{2}+a_i+a_j \\  n+\frac{1}{2}+a_i+a_j & n+a_i+a_j  \\ \end{array} & \begin{array}{cc} 0 & n+\frac{1}{2} + a_i - a_j \\  n+\frac{1}{2} + a_i-a_j & 0 \end{array} \end{array}\right)\,, 
\end{equation}  
\begin{equation}
    \label{eq:so(2n)A-}
    [Q^{\rm diag},A^{-}_\mu] = \left( \begin{array}{c|c} \begin{array}{cc}  0 &  n -a_i+a_j \\  n-a_i+a_j & 0 \end{array} & \begin{array}{cc}0 & n-a_i-a_j \\ n-a_i-a_j & 0 \end{array} \\ \hline \begin{array}{cc}  0 & n+a_i+a_j \\  n+a_i+a_j & 0  \\ \end{array} & \begin{array}{cc} 0 & n + a_i - a_j \\  n + a_i-a_j & 0 \end{array} \end{array}\right)\,.
\end{equation}  
Given these spectra, we notice that the contribution from $A^{+}_{\mu}$ and $A^{-}_{\mu}$ are not the same under the exchange of $\mathcal{F}^{+}$ and $\mathcal{F}^{-}$, which is consistent with the fact that they contain a different number of fields. To compute the potential we need to look at the substructure of Eqs.~\eqref{eq:so(2n)A+}-\eqref{eq:so(2n)A-}.
For $A^{+}_{\mu}$, the diagonal blocks give half of an $\SU(N)$ contribution with one type of VEV, while the off-diagonal blocks contribute schematically as:
\begin{equation}
     \frac{1}{2} \left. \mathcal{V}_{Adj}\right|_{\SU(N)}+[\text{diagonal elements}]\,.
\end{equation}
 For $A^{-}_{\mu}$ each of the diagonal/off-diagonal blocks will give half a contribution of $\SU(N)$ with one parity flipped.  The total potential then becomes
\begin{equation}
    \begin{split}
       \left.\mathcal{V}^{{\rm I+II}}_{Adj} \right|_{\SO(2N)} = & \sum_{i,j=1}^{n_{pq}}\left[\mathcal{F}^{+}(a_i+a_j)+\mathcal{F}^{+}(a_i-a_j)\right]+\sum_{i,j=1}^{n_{pq}}\left[\mathcal{F}^{-}(a_i+a_j)+\mathcal{F}^{-}(a_i-a_j)\right]+\\
    & +2\vert p-q\vert\sum_{i,j=1}^{n_{pq}}\mathcal{F}^{+}(a_i)+2\vert p-q\vert\sum_{i,j=1}^{n_{pq}}\mathcal{F}^{-}(a_i)+\sum_{i,j=1}^{n_{pq}}\mathcal{F}^{+}(a_i+a_j)\,. 
    \end{split}
\end{equation}

%%%%%%%%%%%%%%%%%%%%%%%%%%%%%%%%%
\subsubsection{Case II+II}
%%%%%%%%%%%%%%%%%%%%%%%%%%%%%%%%%

The last case we can look at is when both parities are of the type II. Once again, we split the components of the gauge fields depending on the $X^+$ and $X^-$ generators. The parity assignments are
\begin{equation}
        (P_1,P_2)(A_\mu^+) = \left( \begin{array}{c|c} \begin{array}{cc} {(+,+)} & (+,-) \\ (+,-) & {(+,+)} \end{array} & \begin{array}{cc} {(+,+)} & (+,-) \\ (+,-) & {(+,+)} \end{array} \\ \hline \begin{array}{cc} {(+,+)} & (+,-) \\ (+,-) & {(+,+)} \end{array} & \begin{array}{cc} {(+,+)} & (+,-) \\ (+,-) & {(+,+)} \end{array} \end{array}\right)\,,
     \label{eq:A+ parities}
\end{equation}
\begin{equation}
        (P_1,P_2)(A_\mu^-) = \left( \begin{array}{c|c} \begin{array}{cc} {(-,-)} & (-,+) \\ (-,+) & {(-,-)} \end{array} & \begin{array}{cc} {(-,-)} & (-,+) \\ (-,+) & {(-,-)} \end{array} \\ \hline \begin{array}{cc} {(-,-)} & (-,+) \\ (-,+) & {(-,-)} \end{array} & \begin{array}{cc} {(-,-)} & (-,+) \\ (-,+) & {(-,-)} \end{array} \end{array}\right)\,. 
         \label{eq:A- parities}
\end{equation}
The gauge-scalar zero modes can also be found in $A_{\mu}^-$, but this time they are in the antisymmetric representations of $\U(p)$ and $\U(q)$. Without loss of generality, we can define a VEV for $A_5$ in the following way:
\begin{equation}
    Q = \left( \begin{array}{cc|cc} 
     \mbox{diag}[a_i] & 0 &0&0\\
    0 & \mbox{diag}[b_j] &0&0  \\\hline 0&0& \mbox{diag}[a_i] & 0\\
    0&0& 0 & \mbox{diag}[b_j]  
    \\
    \end{array}\right)  \otimes \begin{pmatrix}
        0 & 1 \\
        -1 & 0 
        \end{pmatrix} \,.
\end{equation}
This charge can be diagonalised in a similar fashion. We can compute again the spectrum for $A_{\mu}^{\pm}$. As with the case with parities I+II, some contributions in $A_{\mu}^+$ are not found in $A_{\mu}^-$ as the number of degrees of freedom is not the same. As we are working with antisymmetric representations, there is no off-diagonal blocks. We then get the following potential:
\begin{equation}
\begin{split}
    \left.\mathcal{V}^{{\rm II+II}}_{Adj}\right|_{\SO(2N)} & = \frac{3}{2} \sum_{i,k=1}^p \left[ \mathcal{F}^+(a_i+a_k)+\mathcal{F}^+(a_i-a_k) \right] + \frac{3}{2} \sum^q_{j,l=1} \left[\mathcal{F}^+(b_j+b_l)+\mathcal{F}^+(b_j-b_l) \right] \\
    & + 4 \sum_i^p \sum_j^q \left[ \mathcal{F}^-(a_i+b_j) +\mathcal{F}^-(a_i-b_j) \right]\,.
\end{split}
\end{equation}

%%%%%%%%%%%%%%%%%%%%%%%%%%%%%%%%%%%%%%%%%%%%%%%%%%%%%%%%%%%%%%%%
\section{Stable $\SO(10)/\Sp(10) \to \SU(3) \times \SU(2) \times \U(1)^2$} \label{app:SO(10)}
%%%%%%%%%%%%%%%%%%%%%%%%%%%%%%%%%%%%%%%%%%%%%%%%%%%%%%%%%%%%%%%%

As we have seen, both $\Sp(10)$ and $\SO(10)$ can be broken to $\U(3) \times \U(2)$ by use of a combination of parity types I and II. The parity of type II breaks the group to $\SU(5) \times \U(1)$, while the additional breaking of $\SU(5)$ can be obtained via a misalignment in the parity of type I, encoded by an $\SU(5)$ parity matrix $P_5 = \mbox{diag} (+1,+1,+1,-1,-1)$. In both cases, therefore, it will be convenient to characterise the representations in terms of $\SU(5)$ components and their parities.

%%%%%%%%%%%%%%%%%%%%%%%%%%%%%%
\subsection{$\Sp(10)$ model}
%%%%%%%%%%%%%%%%%%%%%%%%%%%%%%

The two most promising representations are the fundamental $\bf 10$ and the antisymmetric $\bf 45$. They decompose under $\SU(5) \times \U(1)$ as
\begin{equation}
    {\bf 10} \to 5_1 \oplus \overline{5}_{-1}\,, \quad {\bf 45} \to 10_2 \oplus \overline{10}_{-2} \oplus 24_0 \oplus 1_0\,.
\end{equation}
Hence, the SM can be embedded inside the $\SU(5)$ group, broken by $P_5$.
The parities of the components under the SM can, therefore, be expressed in terms of the diagonal entries of $P_5$. For the fundamental $\bf 10$ representation,
\begin{equation}
    5_1 \to (P_5, +)\,, \qquad \overline{5}_{-1} \to (-P_5, -)\,.
\end{equation}
These parities match those of the minimal $\SU(5)$ aGUT \cite{Cacciapaglia:2020qky}, therefore, the $\bf 10$ representation can be arranged to contain the lepton doublets and the right-handed down quarks. However, for the $\bf 45$ representation, we obtain
\begin{equation}
    10_2 \to (P_5, +)\,, \quad \overline{10}_2 \to (P_5, +)\,, \quad 25_0\oplus 1_0 \to (-P_5, -)\,.
\end{equation}
As the two $10$ have the same parities, one cannot use them to embed the remaining SM fermions. For this reason, the $\Sp(10)$ case is ruled out.

%%%%%%%%%%%%%%%%%%%%%%%%%%%%%%
\subsection{$\SO(10)$ model}
%%%%%%%%%%%%%%%%%%%%%%%%%%%%%%

A similar construction can be applied to the $\SO(10)$ bulk gauge group. As for $\Sp(10)$, the fundamental and antisymmetric representations do not contain the SM fermions for very similar reasons. However, orthogonal groups also allow for spinorial representations, which are in fact used in the traditional $\SO(10)$ GUTs. Contrary to the other representations, the spinorial $\bf 16$ representation of $\SO(10)$ is complex. Note that this orbifold has been considered in Ref. \cite{Haba_so10}.

As for the previous case, we can decompose the spinorial in terms of $\SU(5)$ components
\begin{equation}
    {\bf 16} \to 10_1 \oplus \overline{5}_{-3} \oplus 1_5\,,
\end{equation}
with parities,
\begin{equation}
    10_1 \to (-P_5, -)\,, \quad \overline{5}_{-3} \to (P_5, +)\,, \quad 1_5 \to (-,+)\,.
\end{equation}
Comparing with the analysis in Ref. \cite{Cacciapaglia:2020qky}, one can immediately see that the relative parities between the $10$ and $\overline{5}$ do not allow to embed the SM fermions. Hence, this $\SO(10)$ model needs to be discarded as well.

\bibliographystyle{utphys}
 
\bibliography{bib}

\end{document}